**Multimode optomechanical dynamics in a cavity with avoided crossings**

D. Lee[1], M. Underwood[1], D. Mason[1], A. B. Shkarin[1], S. W. Hoch[1] and J. G. E. Harris[1,2*]

[1] Department of Physics, Yale University, New Haven, CT, 06511, USA

[2] Department of Applied Physics, Yale University, New Haven, CT, 06511, USA

*e-mail: jack.harris@yale.edu

Cavity optomechanics offers powerful methods for controlling optical fields and mechanical motion. A number of proposals have predicted that this control can be extended considerably in devices where multiple cavity modes couple to each other via the motion of a single mechanical oscillator. Here we study the dynamical properties of such a multimode optomechanical device, in which the coupling between cavity modes results from mechanically-induced avoided crossings in the cavity's spectrum. Near the avoided crossings we find that the optical spring shows distinct features that arise from the interaction between cavity modes. Precisely at an avoided crossing, we show that the particular form of the optical spring provides a classical analog of a quantum-nondemolition measurement of the intracavity photon number. The mechanical oscillator's Brownian motion, an important source of noise in these measurements, is minimized by operating the device at cryogenic temperature (500 mK).

Optomechanical systems are typically modelled as a single cavity mode whose eigenfrequency is proportional to the displacement of a mechanical oscillator.[1] This "single-mode" model of optomechanics gives an accurate description of devices in which there is a clear separation of frequencies (e.g., between the mechanical frequency and the cavity mode spacings), and when only a single cavity mode is strongly driven.[2] Single-mode optomechanical devices have been used to realize a number of goals in recent years, including demonstrations of quantum effects associated with Gaussian states of the cavity field and/or the mechanical oscillator.[3,4,5,6,7,8,9,10,11,12,13,14]

For some optomechanical devices the single-mode description breaks down and more complex behavior can occur. In particular, devices in which multiple cavity modes couple to each other via the oscillator's motion are predicted to offer novel means for controlling and measuring both mechanical motion and electromagnetic fields.[15,16,17,18,19,20,21,22,23] Such a mechanical coupling between cavity modes can be produced by applying strong coherent drives to the modes (in which case adiabatic elimination of the mechanical degree of freedom results in an effective coupling between the drives' sidebands).[22,23] This approach can be realized in a very wide range of optomechanical systems, since most cavities possess several modes that can be driven strongly, and whose eigenfrequencies depend upon the oscillator's displacement. Recent experiments have used this approach[24,25] (or a related approach that combines strong drives with a piezoelectric material[26]) to transfer modulation sidebands between different wavelengths, including from microwave to near-infrared.

A different method for mechanically coupling cavity modes (and one which does not require multiple strong drives) is to employ devices in which the cavity's eigenmodes (rather than eigenfrequencies) depend strongly upon the oscillator's displacement. This situation occurs when the oscillator's displacement causes crossings in the cavity's spectrum: these crossings are typically avoided (owing to broken symmetries within the device),[27,28,29] and in the vicinity of each avoided crossing the cavity's eigenbasis depends strongly upon the oscillator's displacement.[30,29] Theoretical studies of the resulting coupling show that it can offer improved performance over single-mode devices, e.g. in producing squeezed states of the mechanical oscillator and optical field.[19,20,21] Perhaps more importantly, the multimode coupling associated with avoided crossings offers capabilities that are fundamentally distinct from those of single mode devices, with applications in macroscopic matter-wave interferometry[18] and measuring the phonon statistics of a driven mechanical oscillator.[15,16]

Avoided crossings are not a generic feature in optomechanical systems, but have been demonstrated in devices based on the membrane-in-the-middle design,[27,29,31,32] ultracold atoms,[33] and whispering gallery mode resonators.[34,35] To date, measurements of these systems have mostly focused on static spectroscopy of the cavity modes (i.e., to determine the parameters of the avoided crossings).[27,28,29,31,32,33,34,35] However the utility of the avoided crossings arises from their dynamical effects, which have received much more limited experimental study.[33,34,35]

Here we address three outstanding issues related to multimode optomechanical devices based on cavities with avoided crossings. First, we describe thorough measurements of the optomechanical dynamics in the vicinity of avoided crossings. Far from the crossings, we find behavior that is dominated by the conventional dynamical back action[1] of the laser driving the cavity; in contrast, near the crossings the behavior is dominated by the elastic energy stored by the intracavity light. Second, we exploit the elasticity of the intracavity light at the crossings to demonstrate a classical analog of a quantum non-demolition (QND) measurement of the cavity's photon number. Third, the device is operated at temperature $T = 500$ mK, which minimizes the impact of thermomechanical noise, and should aid in future work directed at observing quantum effects in multimode optomechanical systems.

These results complement earlier studies of classical multimode dynamics in different systems, for example in purely mechanical devices,[36,37] purely electromagnetic devices,[38] and devices in which multiple mechanical modes couple via a single electromagnetic mode.[39,40,41]

**Experimental setup**

The experimental setup is shown in Fig 1a. It consists of a $Si_3N_4$ membrane (1 mm × 1mm × 50 nm) placed inside a Fabry-Perot optical cavity and cooled by a $^3$He cryostat to $T = 500$ mK. The cavity finesse $F = 4,000$ (linewidth $\kappa/2\pi = 1$ MHz), and the membrane's fundamental mode resonates at $\omega_m/2\pi = 354.6$ kHz with quality factor $Q = 100,000$. Laser light with wavelength $\lambda = 1064$ nm enters the cryostat via an optical fiber. This light is coupled from the fiber to the cavity via cryogenic free-space optics which are aligned *in situ* using piezoelectric motors. Similar motors are used to adjust the membrane's position, tip, and tilt within the cavity. An additional piezoelectric element allows for fine displacement of the membrane along the cavity axis, and for excitation of the membrane's vibrational modes.

Two lasers are used to address two cavity modes that are separated by 8.13 GHz (roughly twice the free spectral range). The first laser is the "probe" beam; it is locked to the cavity and detects the membrane's motion via a heterodyne scheme. The second laser is the "control" beam, and is locked to the probe beam with a controllable frequency offset. This control beam produces the multimode optomechanical interactions that are the main focus of this paper. Additional information about the setup is provided in the Supplemental Information.

**Static spectroscopy**

Figures 1b and c show cavity reflection spectra measured separately by the probe beam (upper plots) and the control beam (lower plots). In each case the reflection was recorded as a function of laser detuning and the membrane's static displacement $z_{dis}$. The brightest curve corresponds to the TEM$_{00}$ mode ('singlet'), while the slightly dimmer curves correspond to the TEM$_{\{20,11,02\}}$ ('triplet') modes. The triplet modes are nearly degenerate, but can be resolved in the closer view shown in Fig. 1c.

The longitudinal order of the singlet mode differs by one from that of the triplet modes; as a result their resonance frequencies $\omega_{cav}$ undergo roughly opposite detuning as a function of $z_{dis}$,[28] and so appear to cross each other near $z_{dis} = 0$ nm and $z_{dis} = -160$ nm. A closer view of the apparent crossing near $z_{dis} = 0$ nm shows that two of the triplet modes avoid the singlet mode (Fig. 1d).[29] The optomechanical dynamics that occur near these avoided crossings is the main focus of this paper.

Because the probe and control beams address modes with slightly different wavelength, the avoided crossings for the two beams occur at different values of $z_{dis}$. This makes it possible to position the membrane so that the probe beam addresses a mode that is *not* part of an avoided crossing (and so simply provides an efficient readout of the membrane's oscillatory motion $z_{osc}(t)$) while the control beam addresses modes that undergo an avoided crossing (thereby producing multimode optomechanical coupling). Such a position is indicated in Fig 1c as a dashed white line, which we define as $z_{dis} = 0$ nm.

**Optomechanical dynamics near avoided crossings**

To demonstrate the impact of the avoided crossings on the membrane's motion, we first position the membrane at $z_{dis} = 0$ nm where the detuning of the modes addressed by the control beam is quadratic to lowest order, i.e. $\omega_{cav} \propto z_{osc}^2$. In this case, each intracavity photon is predicted[27,42] to produce an optical spring that shifts $\omega_m$ by an amount $g_2 = \omega_{cav}'' z_{ZP}^2$ (the primes indicate differentiation with respect to $z_{osc}$, and $z_{ZP}$ is the amplitude of the membrane's zero-point motion). Fig. 1e plots the power spectral density of the membrane's Brownian motion (recorded by the probe beam) as the control beam's detuning $\Delta$ is varied.

This data shows the two qualitative features of quadratic coupling. First, the change in the membrane's resonance frequency $\delta\omega_m$ is proportional to the number of intracavity photons (i.e., $\delta\omega_m(\Delta)$ has even symmetry about each cavity resonance with an approximately Lorentzian shape). Second, the sign of $\delta\omega_m$ is set by the sign of $\omega_{cav}''$ (i.e., positive when the laser is tuned to the higher-frequency cavity mode, and negative when the laser is tuned to the lower-frequency mode). In contrast, for conventional single-mode optomechanics (in which the detuning is linear: $\omega_{cav} \propto z_{osc}$) $\delta\omega_m(\Delta)$ has odd symmetry about a cavity resonance, and its sign is the same regardless of which cavity mode is excited by the laser.[1]

To make a more quantitative comparison with theory, we use multimode optomechanics theory[42] to calculate the cavity reflection, optical spring, and optical damping in the presence of avoided crossings (see Methods and Supplemental Information for more details). The majority of the parameters in this theory are determined by fitting the cavity's static spectrum to expressions that include three cavity modes (fig. 1d shows a comparison of the measured (left) and fitted (right) reflection). To determine the remaining parameters, and to test the predictions of this model with respect to dynamical behavior, we measured the membrane's Brownian motion at several values of $z_{dis}$ between -1 nm and +1.25 nm. At each value of $z_{dis}$, the control beam detuning $\Delta$ was varied over a range that included both of the cavity modes participating in the avoided crossing. For each value of $\Delta$, the membrane's resonance frequency $\omega_m$ and mechanical damping rate $\gamma_m$ were determined by fitting the Brownian motion spectrum. Figure 2 shows the changes in these quantities (i.e., the optical spring $\delta\omega_m$ and the optical damping $\delta\gamma_m$) as a function of $\Delta$ for each value of $z_{dis}$.

When the membrane is furthest from the avoided crossing (i.e., for the uppermost and lowermost curves in Fig. 2), the features in $\delta\omega_m$ and $\delta\gamma_m$ show odd symmetry about the cavity resonances (which are indicated by dashed lines), consistent with conventional single-mode optomechanics and linear coupling. As $z_{\text{dis}}$ approaches 0 nm, the features in $\delta\omega_m$ and $\delta\gamma_m$ decrease in size, consistent with the decreasing slope of the cavity detuning near the avoided crossing. Precisely at the avoided crossing (olive-colored data in Fig. 2) the odd-symmetry feature in $\delta\omega_m$ is completely absent, and is replaced by an even-symmetry feature (as discussed above in the context of Fig. 1e).

The solid lines in Fig. 2 are calculated from the model described in Methods. These calculations use the parameters determined from the cavity's static spectrum (Fig. 1d), as well as three additional fit parameters. A complete description of the fitting process is given in the Supplementary Information. The agreement between the data and the fits in Fig. 2 indicates that multimode optomechanics theory provides an accurate description of this system, particularly in the vicinity of multiple avoided crossings between cavity modes.

Figure 3 shows similar measurements, but carried out at fixed $z_{\text{dis}} \approx 0$ nm as a function of the control beam power $P_{\text{in}}$. The data are plotted along with the predictions of the model. These predictions use the parameter values taken from the fits in Fig. 2, except for $z_{\text{dis}}$ and $P_{\text{in}}$ which are used as fit parameters (the fit values of $z_{\text{dis}}$ and $P_{\text{in}}$ agree well with independent measurements, as described in the Supplementary Information). Figure 3 shows clearly that when $z_{\text{dis}} \approx 0$ nm, the feature in $\delta\omega_m$ has even symmetry at each cavity resonance while the feature in $\delta\gamma_m$ has odd symmetry, in agreement with theory.

Previous measurements of static reflection spectra at room temperature showed that it is possible to tune the avoided crossings by adjusting the membrane's tilt relative to the cavity axis, and its position along the cavity axis[29,32] (see also Ref. [35]). To illustrate this capability at $T = 500$ mK and to demonstrate its impact on the optomechanical dynamics, Figures 4a and b show cavity spectra for two different membrane alignments. When the membrane is positioned near the cavity waist with nominally zero tilt (Fig. 4a), only one of the triplet modes forms an avoided crossing with the singlet mode. After translating the membrane by -15 μm along the cavity axis and tilting it by 0.3 mrad (Fig. 4b), two of the triplet modes form avoided crossings.

Figure 4c shows measurements of $\delta\omega_m(\Delta)$ for each of the three avoided crossings in Figs. 4a and b. For each measurement, $z_{\text{dis}}$ was set so that the membrane was at the avoided crossing. The solid lines are fits to the same model as the previous figures. As the avoided crossing gap is decreased, the peaks in $\delta\omega_m$ move closer together and grow larger, reflecting the increase in $\omega''_{\text{cav}}$. For the uppermost trace, the gap at the avoided crossing is no longer substantially larger than $\kappa$, and the two peaks begin to merge. See Table S2 in the Supplementary Information for a full description of the fit results.

**Classical analog of a photon QND measurement**

Proposals for realizing a QND measurement of the membrane's phonon number or the cavity's photon number make use of the fact that at an avoided crossing, a change in the number of quanta in one oscillator (optical or mechanical) shifts the frequency of the other oscillator by $g_2$. Fully realizing these proposals and using them to detect individual quantum jumps requires single-quantum strong coupling,[30] which has not been achieved for optomechanical devices to date. Instead, we demonstrate a classical analog of such a measurement by using the membrane's resonance frequency $\omega_m$ to monitor classical fluctuations of the intracavity laser power.

These fluctuations are produced by modulating the power of the control laser with frequency 75 Hz and depth 0.77. At the same time, the membrane's fundamental mode is driven (using the piezo element) in a phase-locked loop (PLL). The PLL ensures that the frequency of the piezo drive tracks fluctuations in $\omega_m$, and the PLL error signal provides a record of these fluctuations (see Supplementary Information for details). Figure 5a shows $S_{ff}$, the spectrum of these fluctuations, when the membrane is positioned at an avoided crossing ($z_{dis} = 0$ nm in Fig. 4a) and the control beam is tuned to the cavity resonance ($\Delta = 0$). The peak in $S_{ff}$ at 75 Hz reflects the response of $\omega_m$ to the laser's modulation.

Figure 5b and c show $A_\omega$, the amplitude of the 75 Hz modulation of $\omega_m$ as a function of $\Delta$. In Fig. 5b, $A_\omega(\Delta)$ is measured with $z_{dis} = 0$ nm (i.e., at an avoided crossing). The maximum value of $A_\omega$ occurs at $\Delta = 0$, as expected for quadratic coupling. In contrast, Fig. 5c shows $A_\omega(\Delta)$ measured with $z_{dis} = 3$ nm (i.e., far from an avoided crossing). In this case, $A_\omega$ has a minimum at $\Delta = 0$, as expected for linear coupling. The solid lines in Figs. 5b and c are fits to the same model as in the other figures. We emphasize that the important difference between the quadratic and linear coupling is not the specific form of $A_\omega(\Delta)$ (although measuring $A_\omega(\Delta)$ does provide a simple practical means for distinguishing them), but rather the different physical mechanisms by which the two couplings produce an optical spring. Specifically, the optical spring associated with linear coupling in single-mode devices arises from the leakage of light into and out of the cavity with each oscillation of the membrane.[1] In contrast, the optical spring associated with quadratic coupling results from the elastic energy stored in the intracavity field.[34] This distinction underlies a number of the proposed applications of these avoided crossings.[15,16,18]

**Summary and outlook**

In summary, we have measured the dynamics of an optomechanical device in which multiple cavity modes are coupled by the motion of a single mechanical oscillator. We find that avoided crossings between the cavity modes result in an optical spring that differs substantially from conventional, single-mode optomechanical devices. These results are in quantitative agreement with a classical theory of the device's linear dynamics. This agreement, along with the demonstration of this device's *in situ* tunability and cryogenic operation, are important steps towards studying the nonlinear and quantum regimes of multimode optomechanical devices. In particular, by improving this device's cavity finesse and mechanical quality factor (as demonstrated in Ref. [14]), it should be possible to exploit multimode effects to efficiently produce squeezed states of the mechanical oscillator and optical field,[19,20,21] transfer states between cavity modes,[17] initialize macroscopic matter-wave interferometers,[18] and measure the quantum statistics of a driven mechanical oscillator.[15,16]

**Methods**

Following the description in Ref. [42], we represent the cavity field as a superposition of basis modes, which we take to be the cavity's eigenmodes when the membrane is far from the avoided crossings. The amplitudes of these modes, $a_n$, are the cavity's degrees of freedom. The membrane couples these modes and detunes them by an amount that depends upon $z_{\text{dis}}$ and $z_{\text{osc}}$ (here $z_{\text{dis}}$ is the uniform translation of the membrane chip, and $z_{\text{osc}}$ is the instantaneous displacement associated with the membrane's oscillatory motion). For the small range of motion considered here, we assume this detuning is linear in both $z_{\text{dis}}$ and $z_{\text{osc}}$. These effects can be incorporated into the usual optomechanical equation of motion via the Hamiltonian $H_1 = \vec{a}^\dagger M \vec{a} + \hbar \omega_{\text{m}}^{(0)} b^\dagger b$, where the components of the vector $\vec{a}$ are the mode amplitudes $a_n$, $b$ is the amplitude of the mechanical oscillation, and $M$ is a matrix whose diagonal elements represent the detuning of the cavity modes, and whose off-diagonal terms represent the coupling between modes.[42]

The optomechanical effects associated with avoided crossings emerge from this model even in the simple case of just two optical modes ($n = 1,2$); in this case

$$M = \begin{pmatrix} \omega_c + \omega'_{dis,1}z_{dis} + \omega'_{osc,1}z_{osc} & te^{i\phi} \\ te^{-i\phi} & \omega_c + \omega'_{dis,2}z_{dis} + \omega'_{osc,2}z_{osc} \end{pmatrix} \text{ and } \vec{a} = \begin{pmatrix} a_1 \\ a_2 \end{pmatrix}. \quad (1)$$

This model allows the detuning associated with $z_{dis}$ to have different coefficients ($\omega'_{dis,n}$) from the detuning associated with $z_{osc}$ ($\omega'_{osc,n}$), since the exact location of the cavity mode on the membrane is not known *a priori*. The cavity spectra in Figs. 1b-d correspond to the case where $z_{dis}$ is varied (by scanning the voltage on a small piezoelectric element) while $z_{osc} = 0$ nm. In this case, the two cavity modes would cross at $z_{dis} = 0$ nm, but instead the off-diagonal terms in $M$ produce a gap with magnitude $2t$.

The Supplementary Information provides a more detailed description of this model, and describes how it is used to calculate the optical spring, optical damping, and cavity reflection spectrum. We note that although the restriction to two optical modes (equation (1)) provides an intuitive explanation of most of our data, we use three optical modes ($n = 1,2,3$) for most of the quantitative analysis. Explicit expressions for three optical modes are given in the Supplementary Information; they are straightforward extensions of equation (1) in which $M$ includes two coupling terms ($t_1 e^{i\phi_1}$ and $t_2 e^{i\phi_2}$) corresponding to the two avoided crossings seen in Fig. 1d.

In fitting the cavity spectrum to this model (as in Fig. 1d) there are a large number of fitting parameters; however the fits are highly constrained by the fact that each of the model's parameters corresponds to a prominent feature in the data. For example, the three $\omega'_{dis,n}$ are set by the slopes of the cavity resonances far from the crossings, while the coupling rates $t_1$ and $t_2$ are determined by the magnitudes of the gaps. The coupling phases $\phi_1$ and $\phi_2$ are determined by the amplitudes of the cavity resonances near the crossing. Each mode's $\kappa$ is determined by the width of the resonance far from the crossing, while the input coupling of each mode is determined by the amplitude of the resonance far from the crossing. This analysis of the cavity's static spectrum provides all of the model's parameters except for the three coefficients $\omega'_{osc,n}$. The $\omega'_{osc,n}$ are extracted from fitting the optical spring and optical damping data in Fig 2, as described in the main body of the paper.

**References**

[1] Markus Aspelmeyer, T. J. Kippenberg, and Florian Marquardt, *Cavity optomechanics*, arXiv:1303.0733 (2013).

[2] H. K. Cheung and C. K. Law, *Nonadiabatic optomechanical Hamiltonian of a moving dielectric membrane in a cavity*, Physical Review A **84**, 023812 (2012).

[3] J. D. Teufel, T. Donner, Dale Li, J. W. Harlow, M. S. Allman, K. Cicak, A. J. Sirois, J. D. Whittaker, K. W. Lehnert, R. W. Simmonds, *Sideband Cooling Micromechanical Motion to the Quantum Ground State*, Nature **475**, 359–363 (2011).

[4] Jasper Chan, T. P. Mayer Alegre, Amir H. Safavi-Naeini, Jeff T. Hill, Alex Krause, Simon Gröblacher, Markus Aspelmeyer & Oskar Painter, *Laser cooling of a nanomechanical oscillator into its quantum ground state*, Nature **478**, 89 (2011).

[5] Amir H. Safavi-Naeini, Jasper Chan, Jeff T. Hill, T. P. Mayer Alegre, Alex Krause, and Oskar Painter, *Observation of quantum motion of a nanomechanical resonator*, Physical Review Letters **108**, 033602 (2012).

[6] N. Brahms, T. Botter, S. Schreppler, D. W. C. Brooks, and D. M. Stamper-Kurn, *Optical Detection of the Quantization of Collective Atomic Motion*, Physical Review Letters **108**, 133601 (2012).

[7] T. P. Purdy, R. W. Peterson, and C. A. Regal, *Observation of radiation pressure shot noise on a macroscopic object*, Science **339**, 801 (2013).

[8] Amir H. Safavi-Naeini, Simon Gröblacher, Jeff T. Hill, Jasper Chan, Markus Aspelmeyer, and Oskar Painter, *Squeezed light from a silicon micromechanical resonator*, Nature **500**, 185 (2013).

[9] D.W.C. Brooks, T. Botter, T.P. Purdy, S. Schreppler, N. Brahms, and D.M. Stamper-Kurn, *Non-classical light generated by quantum-noise-driven cavity optomechanics*, Nature **488**, 476 (2012).

[10] T. P. Purdy, P.-L. Yu, R. W. Peterson, N. S. Kampel, and C. A. Regal, *Strong optomechanical squeezing of light*, Physical Review X **3**, 031012 (2013).

[11] T. A. Palomaki, J. D. Teufel, R. W. Simmonds, and K. W. Lehnert, *Entangling Mechanical Motion with Microwave Fields*, Science **342**, 6159 (2013).

[12] A. J. Weinstein, C. U. Lei, E. E. Wollman, J. Suh, A. Metelmann, A. A. Clerk, and K. C. Schwab, "*Observation and interpretation of motional sideband asymmetry in a quantum electro-mechanical device*" ArXiv:1404.3242 (2014).

[13] T. P. Purdy, P.-L. Yu, N. S. Kampel, R. W. Peterson, K. Cicak, R. W. Simmonds, and C. A. Regal, *Optomechanical Raman-Ratio Thermometry*, arXiv:1407.7247

[14] D. Lee, M. Underwood, D. Mason, A. B. Shkarin, K. Børkje, S. M. Girvin, J. G. E. Harris, *Observation of Quantum Motion in a Nanogram-scale Object*, arXiv:1406.7254 (2014)

[15] A. A. Clerk, F. Marquardt, and J. G. E. Harris, *Quantum Measurement of Phonon Shot Noise*, Physical Review Letters **104**, 213603 (2010).

[16] A. A. Clerk, *Full counting statistics of energy fluctuations in a driven quantum resonator*, Physical Review A **84**, 043824 (2011).

[17] Georg Heinrich, J. G. E. Harris, and Florian Marquardt, *Photon Shuttle: Landau-Zener-Stückelberg Dynamics in an Optomechanical System*, Physical Review A **81**, 011801(R) (2010).

[18] Oriol Romero-Isart, *Quantum superposition of massive objects and collapse models*, Physical Review A **84**, 052121 (2011).

[19] H. Seok, L. F. Buchmann, S. Singh, and P. Meystre, *Optically mediated nonlinear quantum optomechanics*, Physical Review A **86**, 063829 (2012).

[20] M. Asjad, G. S. Agarwal, M. S. Kim, P. Tombesi, G. Di Giuseppe, and D. Vitali, *Robust stationary mechanical squeezing in a kicked quadratic optomechanical system*, Physical Review A **89**, 023849 (2013).

[21] Zhi Jiao Deng, Steven J. M. Habraken, Florian Marquardt, *An entanglement rate for continuous variables and its application to a resonant optomechanical multimode setup*, arXiv:1406.7815 (2014).

[22] Amir H. Safavi-Naeini and Oskar Painter, *Proposal for an optomechanical traveling wave phonon–photon translator*, New Journal of Physics **13**, 013017 (2011).

[23] C. A. Regal and K. W. Lehnert, *From cavity electromechanics to cavity optomechanics*, Journal of Physics: Conference Series **264**, 012025 (2012).

[24] Jeff T. Hill, Amir H. Safavi-Naeini, Jasper Chan, and Oskar Painter, *Coherent optical wavelength conversion via cavity-optomechanics*, Nature Communications **3**, 1196 (2012).

[25] R. W. Andrews, R. W. Peterson, T. P. Purdy, K. Cicak, R. W. Simmonds, C. A. Regal and K. W. Lehnert, *Bidirectional and efficient conversion between microwave and optical light*, Nature Physics **10**, 321 (2014).

[26] J. Bochmann, A. Vainsencher, D. D. Awschalom, A. N. Cleland, *Nanomechanical coupling between microwave and optical photons*, Nature Physics **103**, 122602 (2013).

[27] J. D. Thompson, B. M. Zwickl, A. E. Jayich, Florian Marquardt, S. M. Girvin and J. G. E. Harris, *Strong dispersive coupling of a high finesse cavity to micromechanical membrane*, Nature **452**, 72 (2008).

[28] J. C. Sankey, A. M. Jayich, B. M. Zwickl, C. Yang, and J. G. E. Harris, *Improved 'Position Squared' Readout Using Degenerate Cavity Modes*, in Proceedings of the XXI International Conference on Atomic Physics, edited by R. Cote, P. L. Gould, and M. Rozman, (World Scientific, Singapore, 2009).

[29] J. C. Sankey, C. Yang, B. M. Zwickl, A. M. Jayich, and J. G. E. Harris, *Strong and Tunable Nonlinear Optomechanical Coupling in a Low-Loss System*, Nature Physics **6**, 707 (2010).

[30] Haixing Miao, Stefan Danilishin, Thomas Corbitt, and Yanbei Chen, *Standard Quantum Limit for Probing Mechanical Energy Quantization*, Physical Review Letters **103**, 100402 (2009).

[31] N. E. Flowers-Jacobs, S. W. Hoch, J. C. Sankey, A. Kashkanova, A. M. Jayich, C. Deutsch, J. Reichel, J. G. E. Harris, *Fiber-Cavity-Based Optomechanical Device*, Applied Physics Letters **101**, 221109 (2012).

[32] M. Karuza, M. Galassi, C. Biancofiore, C. Molinelli, R. Natali, P. Tombesi, G. DiGiuseppe and D. Vitali, *Tunable linear and quadratic optomechanical coupling for a tilted membrane within an optical cavity: theory and experiment*, Journal of Optics **15**, 025704 (2013).

[33] T. P. Purdy, D. W. C. Brooks, T. Botter, N. Brahms, Z.-Y. Ma, and D. M. Stamper-Kurn, *Tunable Cavity Optomechanics with Ultracold Atoms*, Physical Review Letters **105**, 133602 (2010).

[34] J. T. Hill, Q. Lin, J. Rosenberg, and O. Painter, *Mechanical trapping in a quadratically-coupled optomechanical double disk*, IEEE Conference on Lasers and Electro-Optics (2011).

[35] *Nonlinear optics and wavelength translation via cavity optomechanics*, J. T. Hill, Ph.D. Thesis, California Institute of Technology (2013).

[36] Hajime Okamoto, Adrien Gourgout, Chia-Yuan Chang, Koji Onomitsu, Imran Mahboob, Edward Yi Chang, and Hiroshi Yamaguchi, *Coherent phonon manipulation in coupled mechanical resonators*, Nature Physics **9**, 480 (2013).

[37] T. Faust, J. Rieger, M. J. Seitner, J. P. Kotthaus and E. M.Weig, *Coherent control of a classical nanomechanical two-level system*, Nature Physics **9**, 485 (2013).

[38] R. J. C. Spreeuw, N. J. van Druten, M. W. Beijersbergen, E. R. Eliel, and J. P. Woerdman, *Classical Realization of a Strongly Driven Two-Level System*, Physical Review Letters **65**, 2632 (1990).

[39] Francesco Massel, Sung Un Cho, Juha-Matti Pirkkalainen, Pertti J. Hakonen, Tero T. Heikkilä and Mika A. Sillanpää, *Multimode circuit optomechanics near the quantum limit*, Nature Communications **3**, 987 (2012).

[40] Q. Lin, J. Rosenberg, D. Chang, R. M. Camacho, M. Eichenfield, K. J. Vahala, and O. Painter, *Coherent mixing of mechanical excitations in nano-optomechanical structures*, Nature Photonics **4**, 236 (2010).

[41] A. B. Shkarin, N. E. Flowers-Jacobs, S.W. Hoch, A. D. Kashkanova, C. Deutsch, J. Reichel, and J. G. E. Harris, *Optically Mediated Hybridization between Two Mechanical Modes*, Physical Review Letters **112**, 013602 (2014).

[42] A. M. Jayich, J. C. Sankey, B. M. Zwickl, C. Yang, J. D. Thompson, S. M. Girvin, A. A. Clerk, F. Marquardt, and J. G. E. Harris, *Dispersive optomechanics: a membrane inside a cavity*, New Journal of Physics **10**, 095008 (2008).

**Acknowledgements**

We acknowledge support from AFOSR (grant FA9550-90-1-0484) and NSF (grants 0855455, 0653377, and 1004406). We also thank H. Janssen for technical assistance, and N. Flowers-Jacobs and S. M. Girvin for valuable discussions.

**Author contributions**

D. L., M. U. and D. M. built the experimental set-up, made the measurements and analyzed the data. A. B. S. and S. W. H. developed the theory. J. G. E. H. supervised each phase of the project. All authors contributed to the discussion of the results and the writing of the manuscript.

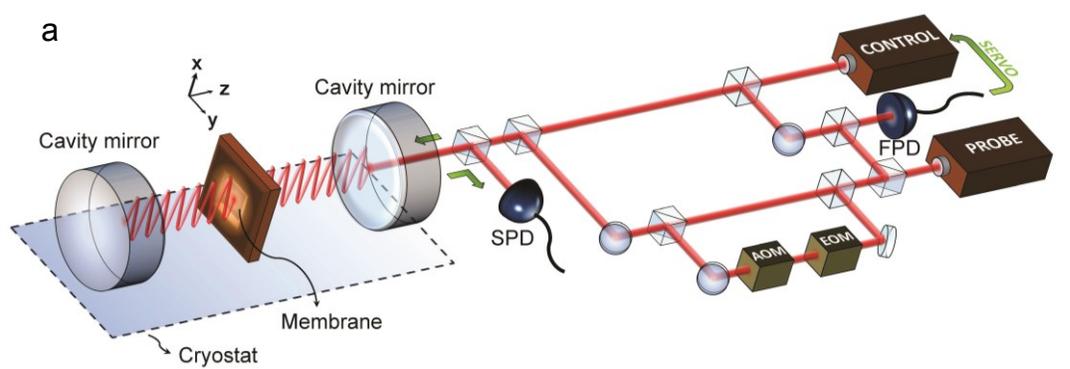
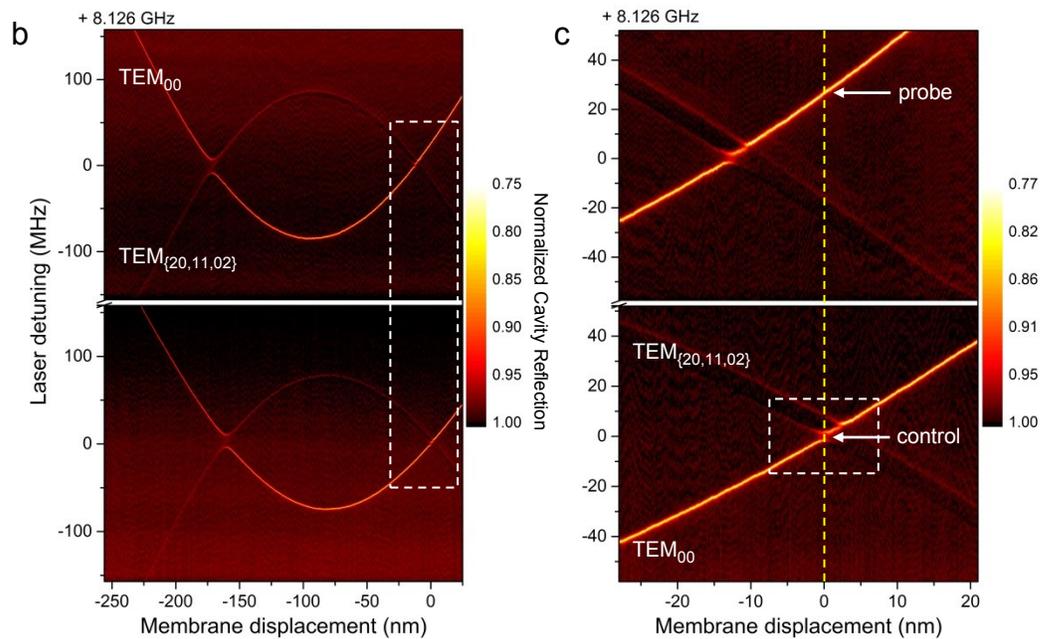
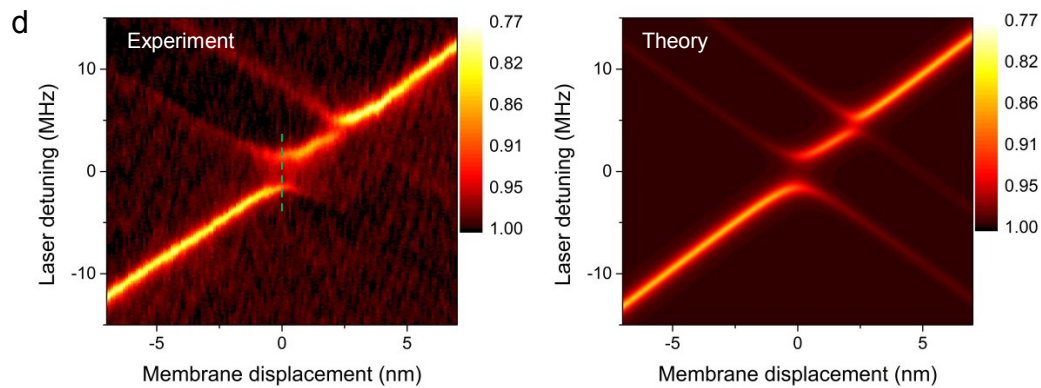
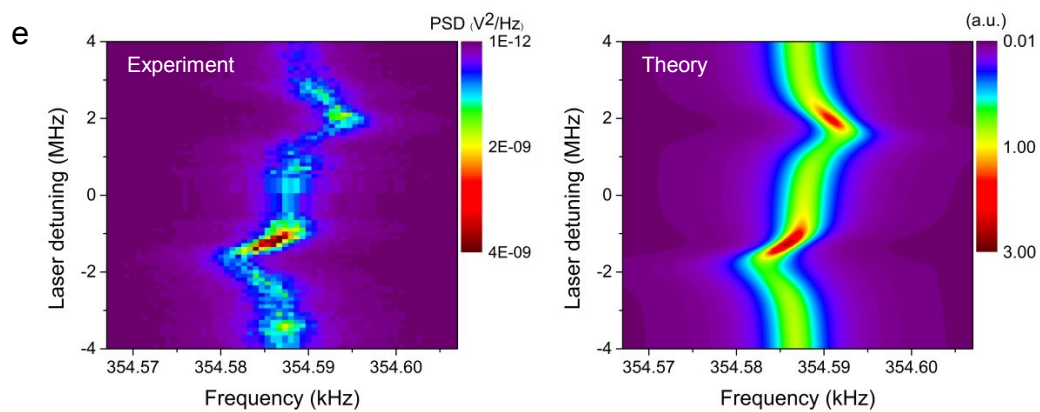

Fig. 1

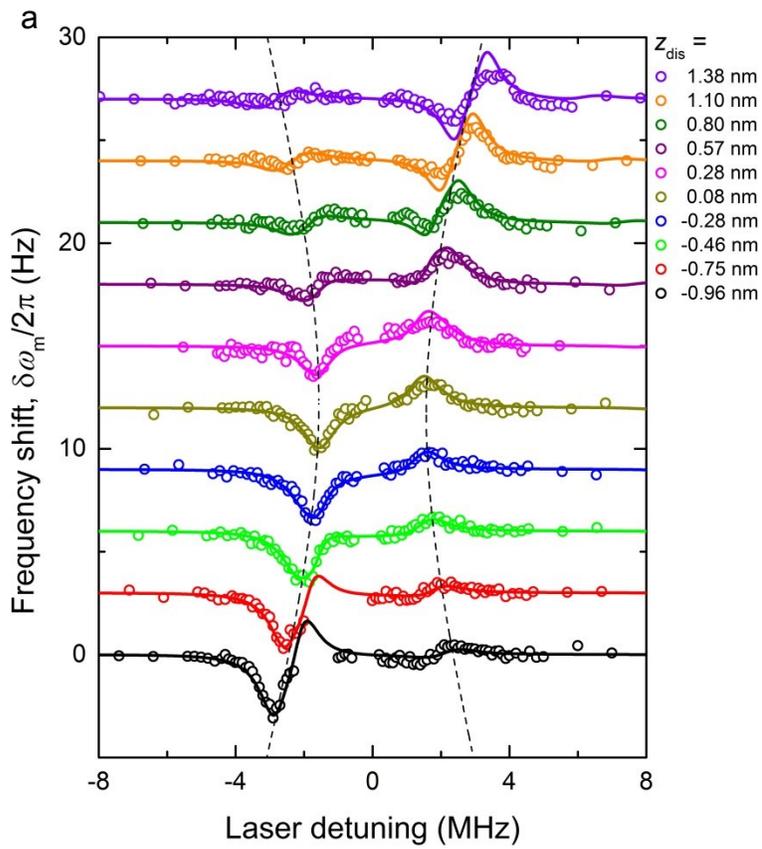 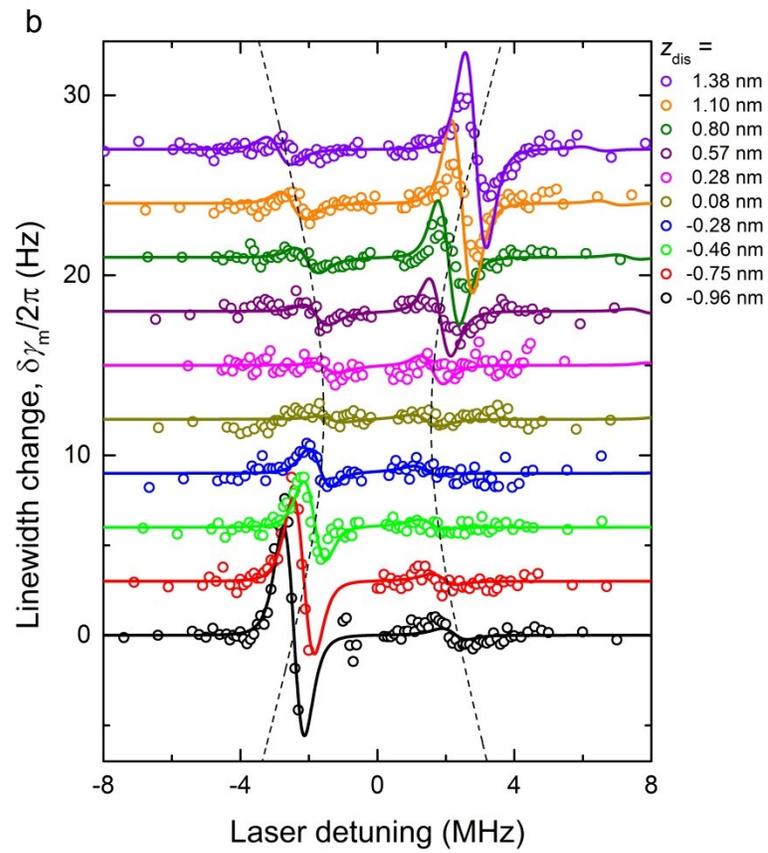

Fig. 2

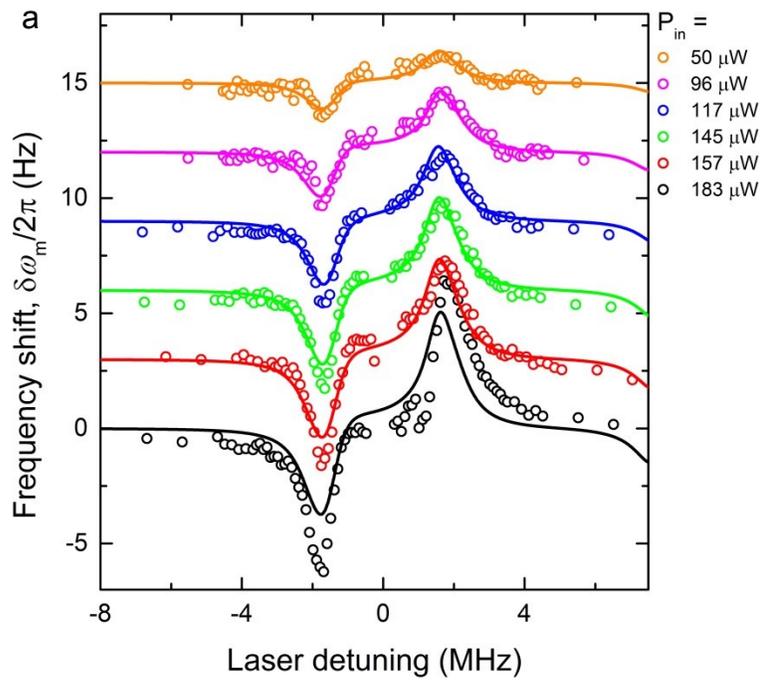 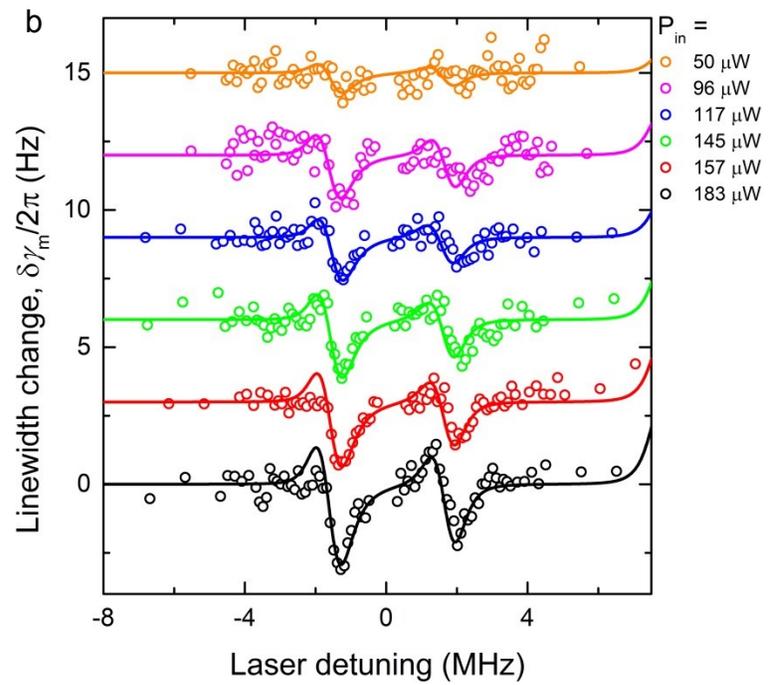

Fig. 3

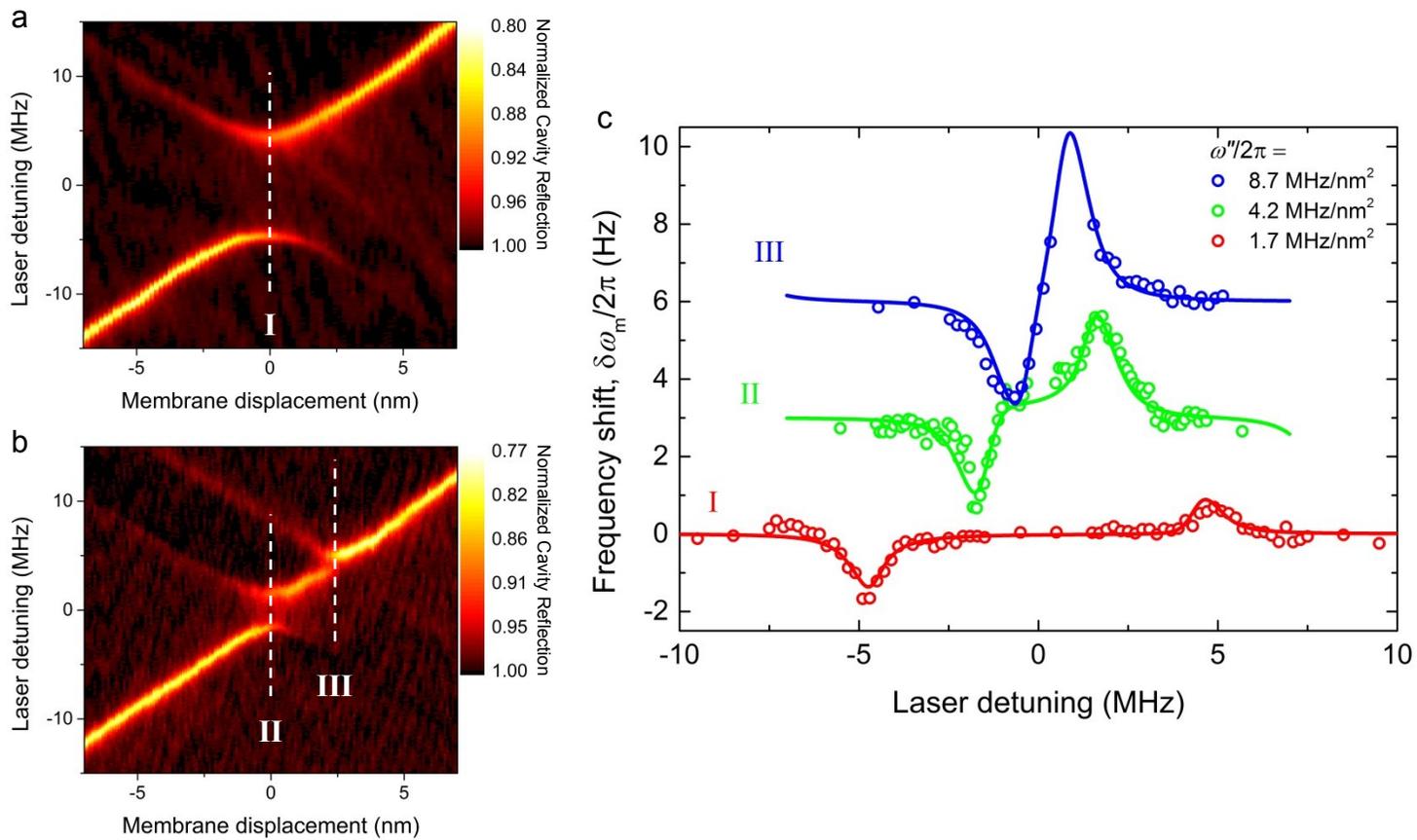

Fig. 4

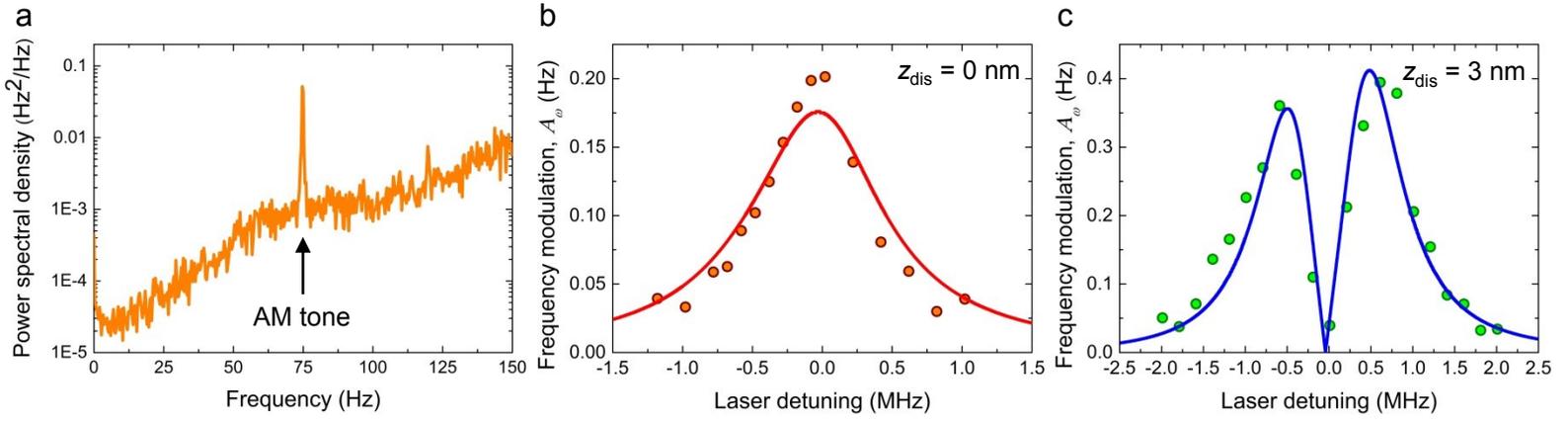

Fig. 5

**Figure captions**

**Figure 1 | System overview and cavity reflection spectroscopy. a**, Schematics of the cryogenic 'membrane-in-the-middle' setup. Two separate lasers ("probe" and "control") address a Fabry-Perot cavity containing a $Si_3N_4$ membrane at $T \sim 500$ mK. Two modulators (AOM and EOM) in the probe beam path allow for Pound-Drever-Hall locking to the cavity and heterodyne detection of the membrane's motion. **b**, Cavity reflectivity, plotted as a function of the membrane's static displacement $z_{dis}$ and laser detuning $\Delta$. The upper and lower plots are measured by the probe and the control lasers, respectively. The cavity's $TEM_{00}$ singlet mode and the $TEM_{\{20,11,02\}}$ triplet modes are visible. **c**, A closer view of the dashed area in b showing avoided crossings between the singlet and triplet modes. The crossings in the modes addressed by the probe beam occur roughly 10 nm away from the crossings in the modes addressed by the control beam. At $z_{dis} = 0$ nm (dashed white line), the probe beam can be used to detect membrane motion while the control beam addresses the avoided crossings. **d**, Zoom-in of the avoided crossings measured with the control beam (left panel) and the calculated cavity spectrum (right panel). **e**, Measured (left panel) and calculated (right panel) power spectral density of the membrane's Brownian motion as a function of control laser detuning $\Delta$ (the range of $\Delta$ is given by the dashed white line in d). For this measurement $z_{dis} = 0$ nm and $P_{in} = 80$ μW. Shifts in the membrane's resonance frequency, consistent with quadratic optomechanical coupling, are visible around the cavity resonances at $\Delta = \pm 1.6$ MHz.

**Figure 2 | Optomechanics near the cavity's avoided crossings. a-b**, Changes in the frequency (a) and linewidth (b) of the membrane's fundamental mode, plotted as a function of control laser detuning $\Delta$ and the membrane's static displacement $z_{dis}$. The avoided crossing occurs at $z_{dis} = 0$ nm. The solid lines are the fits described in Methods and the Supplementary Information. The dashed lines indicate the cavity resonances. For clarity, each curve is shifted vertically by 3 Hz. For large negative values of $z_{dis}$, the lower-frequency cavity mode produces larger optomechanical effects than the higher-frequency cavity due to the fact that it corresponds to the $TEM_{00}$ mode, which is more strongly coupled to the driving laser (as can be seen in Fig. 1d). For large positive values of $z_{dis}$, the situation is reversed.

**Figure 3 | Optomechanics at an avoided crossing. a-b**, Changes in the frequency (a) and linewidth (b) of the membrane's fundamental mode as a function of control laser detuning $\Delta$ and control beam power $P_{in}$. The membrane is nominally at the avoided crossing ($z_{dis} = 0$ nm). $P_{in}$ and $z_{dis}$ are the fit parameters for the theory curves. The fit results for $P_{in}$ are shown in the legend. The fit results for $z_{dis}$ had a mean value of 0.32 nm with a standard deviation of 0.03 nm. For clarity, each curve is shifted vertically by 3 Hz.

**Figure 4 | Optomechanics as the avoided crossings are tuned. a-b**, Cavity reflection spectrum for two different membrane alignments: membrane located at the cavity waist with tilt ~ 0 mrad (a) and translated -15 μm along the cavity axis and tilted 0.3 mrad (b). The three avoided crossings have quadratic coefficients $\omega''_{cav}/2\pi = 1.7$ MHz/nm² (*I*), 4.2 MHz/nm² (*II*), and 8.7 MHz/nm² (*III*). **c**, The membrane's frequency shift measured at the three avoided crossings as a function of the control laser detuning. For each measurement, $P_{in} = 80$ μW. For clarity, each curve is shifted vertically by 3 Hz. See Supplementary

Information for details of the theory and fit results. The data in Fig. 2 and 3 were measured using the crossing with $\omega''_{cav}/2\pi = 4.2$ MHz/nm$^2$.

**Figure 5 | Observing laser fluctuations via quadratic optomechanics. a**, Spectrum of the membrane's resonance frequency, $S_{ff}$ measured using a phase-locked loop. The sharp peak at 75 Hz results from the intensity modulation (modulation depth $\beta = 0.77$) applied to the control beam, which modulates the membrane's frequency via the quadratic optomechanical coupling. **b-c**, The amplitude of the peak in $S_{ff}$, plotted versus control laser detuning at $z_{dis} = 0$ nm (b) and $z_{dis} = 3$ nm (c). The solid lines are fits to the absolute value of the expected optical spring. The fit results are $z_{dis} = -0.14 \pm 0.07$ nm, $\beta = 0.67 \pm 0.15$ for (b) and $z_{dis} = 3.09 \pm 0.01$ nm, $\beta = 0.67 \pm 0.14$ for (c). The quoted errors are statistical fit errors.

# Supplementary information

## Multimode optomechanical dynamics in a cavity with avoided crossings

D. Lee[1], M. Underwood[1], D. Mason[1], A. B. Shkarin[1], S. W. Hoch[1] and J. G. E. Harris[1,2*]

[1] Department of Physics, Yale University, New Haven, Connecticut 06511, USA
[2] Department of Applied Physics, Yale University, New Haven, Connecticut 06511, USA
*e-mail: jack.harris@yale.edu

# 1 Details of experimental setup and methods

### Laser setup

As shown in Fig. 1a, we used two Nd-YAG 1064 nm lasers (Innolight Prometheus) in this experiment. The first laser, which we call the probe laser, is used for cavity locking and for measurement of the membrane's Brownian motion. To make this possible, a portion of the probe beam is sent through an electro-optic modulator (EOM) to apply 15 MHz phase modulation sidebands for the Pound-Drever-Hall (PDH) locking technique. This portion of the beam, (the "PDH beam") also passes through an acousto-optic modulator (AOM) which shifts it by 80 MHz. The frequency-shifted PDH beam is then combined with the unshifted beam which serves as a local oscillator (LO). Both beams are sent into the cryostat to the experimental cavity. Only the PDH beam has the necessary phase modulation sidebands to lock to the cavity, so when the probe laser is "locked", light from the relatively weak PDH beam enters the cavity and interacts with the membrane. The LO beam, which is detuned from the cavity by 80 MHz, promptly reflects off the input mirror of the cavity. When the reflected PDH and LO beams recombine on the signal photodiode (SPD), they produce a beat note at 80 MHz. The membrane's mechanical Brownian motion appears as a phase modulation of this beat note. To observe the beat note, we use a lock-in amplifier to demodulate the signal from the SPD. Typically, the probe beam has about 20 µW PDH power and several hundred µW LO power.

The control laser is nominally identical to the probe laser, except it is detuned in frequency from the probe laser by two cavity free spectral ranges. This frequency offset is produced by mixing a small amount of light from both lasers on the fast photodiode (FPD) shown in Fig. 1a, and comparing the beat note with a reference tone produced by a signal generator. When both lasers are locked to the $TEM_{00}$ mode, they are locked to different longitudinal modes of the cavity, and therefore at a given membrane position, the two lasers may have different couplings to the membrane's motion. This allows us to lock the probe laser to the cavity at a linear point, useful for measurement of the membrane's Brownian motion, and the control laser to the cavity at a quadratic point, useful for producing the effects that we want to study.

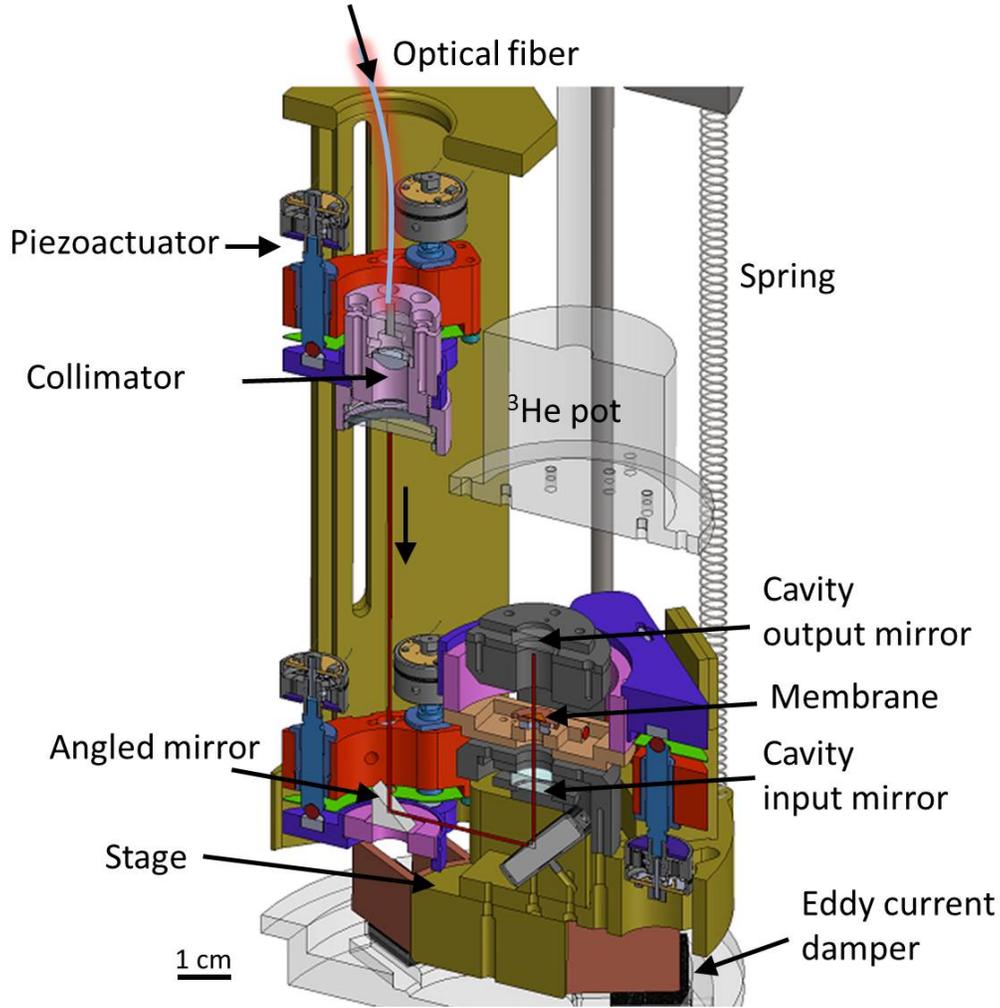

Figure S1: Schematics of experimental cavity setup inside the $^3$He refrigerator.

## Cryostat setup

Light from the two lasers is coupled into the cryostat (Janis Research) through a single-mode optical fiber. Light from the fiber passes through a collimator and then continues in free space, hitting two 45° angled mirrors before reaching the input mirror of the cavity. The fiber collimator and one of the angled mirrors are mounted on custom 1" mirror mounts that can be adjusted *in situ* using commercial piezoelectric actuators (Janssen Precision Engineering, PiezoKnob).

The fiber collimator, mirrors, and cavity are all attached to a titanium stage. The stage is designed to be vibrationally isolated from the outside environment. This is done by suspending the stage on springs inside the cryostat. To reduce oscillatory motion of the stage on the springs, copper eddy current damping fins are attached to the bottom of the stage. Between the fins are strong rare earth magnets. Motion of the stage induces eddy currents in the copper fins, which dissipate the energy as heat. The spring/stage system has a resonance frequency around 2 Hz, and is approximately critically damped by the eddy current dampers. Several hundred flexible gold-coated copper wires (wire diameter of 76 μm) are used for a thermal link between the $^3$He pot ($T \approx 300$ mK) and the stage and membrane. A schematic of the cold experimental cavity setup is shown in Fig. S1.

To provide further vibration isolation, the cryostat itself is attached to a massive aluminum plate, which is mounted on pneumatic air legs. The air legs sit on additional square aluminum plates, which are each supported by four passive vibration reducing feet. The entire system can be enclosed within an acoustic noise reducing "room", consisting of plastic panels coated with sound absorbing foam, to achieve 13 dB of acoustic noise reduction. However, we determined that this level of acoustic isolation was not necessary for the quadratic optomechanics measurement described in this paper, so the acoustic shield was not used in this measurement.

### Phase-locked loop (PLL) measurement

To detect classical laser modulation by way of the optical spring effect, we injected 75 Hz amplitude noise onto the probe laser. This was accomplished by modulating the drive tone of the control beam AOM at 75 Hz with a modulation depth of 0.77.

We then used a piezoelectric element mounted directly beneath the membrane to drive the membrane to an amplitude of 2 nm at its fundamental resonant frequency ($\sim 354.6$ kHz). The 75 Hz amplitude modulation of the control beam causes a 75 Hz modulation of magnitude of the optical spring effect, and therefore modulates the membrane's fundamental frequency at 75 Hz. We used a phase-locked loop (PLL) from a Zurich Instruments HF2LI lock-in amplifier to track the membrane's resonant frequency and detect this 75 Hz modulation, adjusting the frequency of the piezo drive in real time to stay on resonance with the membrane. The output signal of the PLL then contains information about the laser modulation.

# 2 Data analysis and fit results

### Drift subtraction

The membrane's resonant frequency was observed to drift on the order of Hz on a timescale of hours. The amount of drift was sometimes larger than the size of the optical spring shift, which complicated the characterization of the quadratic optomechanical effects. In order to compensate for this drift in our analysis, we always made sure to remeasure the membrane's Brownian motion at selected laser detunings after completing a data run with a given set of parameters. This provided a measurement of the Brownian motion under nominally identical conditions, but at different points in time allowing us to determine the amount by which the membrane's resonant frequency had drifted.

As an example of this process, the membrane's resonant frequency is plotted as a function of laser detuning for 60 µW laser power at $z_{\mathrm{dis}} = 0$ nm (Fig. S2a). This data run took 1 hour and 46 minutes to complete and consisted of a high resolution laser detuning sweep across the avoided crossing (starting at negative detuning), followed by a retaking of selected points in the opposite direction. As can be seen in Fig. S2a, the membrane's mechanical frequency drifts by just under 3 Hz during this time.

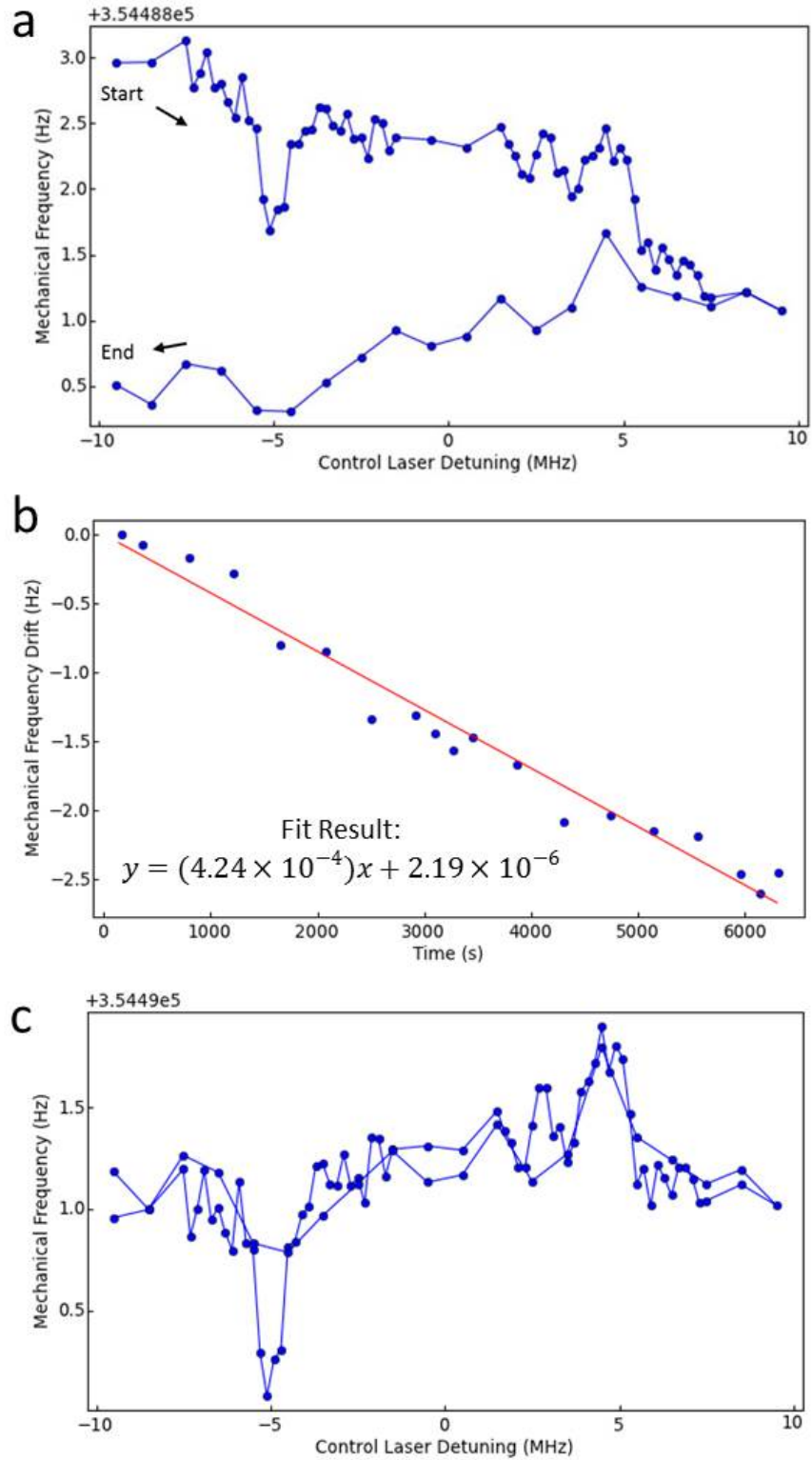

Figure S2: a, Mechanical resonance frequency during forward and backward sweeping of laser detunings. b, Amount of frequency drift as a function of elapsed time. Fit result is shown in the plot. c, Mechanical resonance frequency after the drift correction.

For laser detunings that were measured in both the forward and backward directions, we plotted the difference in the membrane's mechanical frequency as a function of the time passed between the first and second data point at each detuning (Fig. S2b). From the slope of this plot, we determined the rate of membrane resonant frequency drift, and subtracted this drift from the original spring shift data. The corrected data is shown in Fig. S2c. After correction, the data shows a reasonable amount of repeatability despite the time that passed between the forward and backward runs. For actual fitting and data analysis, we discarded the backward run from the post-drift subtraction data.

## System parameters

Our model for predicting optomechanical effects near an avoided crossing depends on a large number of system parameters, including cavity properties, membrane properties, and interaction strengths. When fitting the actual optomechanics data, we would like to minimize the number of free parameters by using independent measurements whenever possible. Our cavity spectrum (as in Fig. S3a) provide an excellent resource for characterizing both the optical properties and some of the interaction strengths in our system.

To completely model the anti-crossing of two optical modes, we need to know the total decay rate of each mode ($\kappa_L$, $\kappa_R$), the decay rate of each mode due to its input mirror ($\kappa_{in,L}$, $\kappa_{in,R}$), the linear coupling between each mode and the membrane's displacement ($\omega'_{dis,L}$, $\omega'_{dis,R}$), and the membrane-mediated coupling rate between the two modes, which we describe as $te^{i\phi}$, with $t$ and $\phi$ real. All of these parameters can be measured from cavity spectroscopy data such as in Fig. S3a.

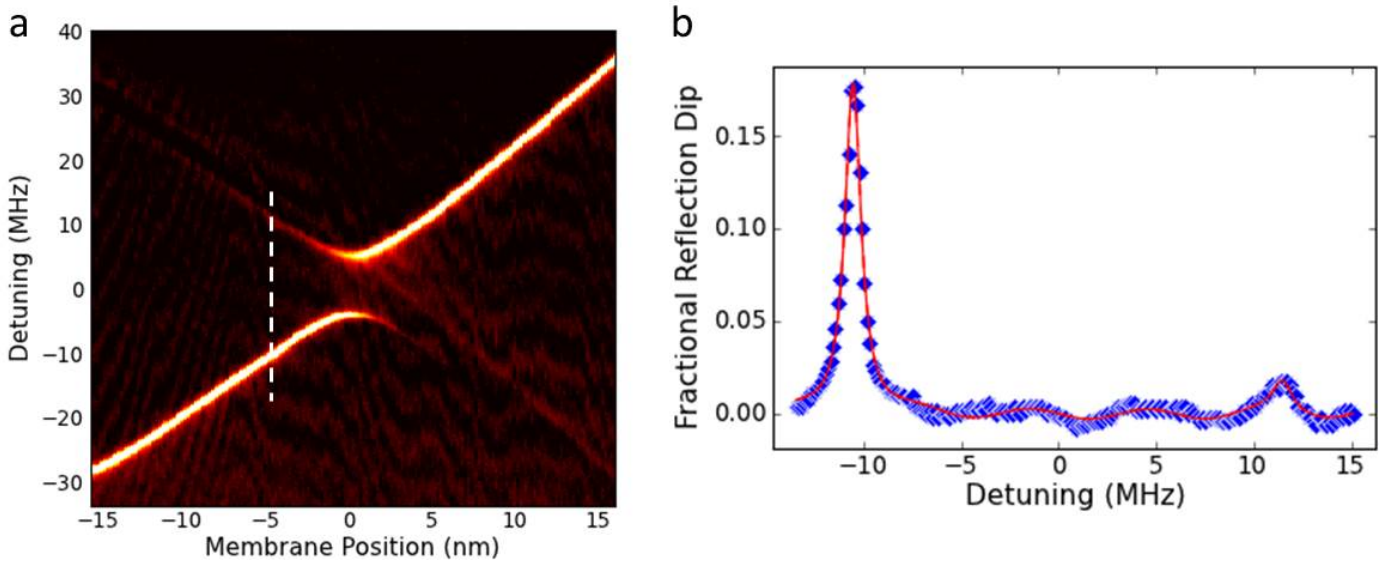

Figure S3: a, Measured cavity spectroscopy showing three triplet modes, one of which couples to the singlet to form an avoided crossing. b, Vertical slice at $z_{dis}$ = -5 nm (dashed line), showing the fractional magnitude of the reflection dips for both the singlet and the triplet. Data is in blue, fit to two Lorentzians on a sinusoidal noise background is in red.

Each vertical slice of the spectrum (e.g. dashed line in Fig. S3a) shows the reflected light intensity measured as the laser driving the cavity is swept over a certain frequency range. Cavity mode resonances appear as Lorentzian peaks whose full width at half maximum (FWHM) is equal to $\kappa$. The 'depth' of the dip provides a measure of $\kappa_{in}$. If we choose a membrane position far from the avoided crossing, then the interaction of the two optical modes can be neglected, and we can make independent measurements of $\kappa$ and $\kappa_{in}$ for both modes. For the two-mode crossing in Fig. S3, we find $\kappa_L/2\pi = 1.0$ MHz, $\kappa_{L,in}/2\pi = 47$ kHz, $\kappa_R/2\pi = 1.3$ MHz, and $\kappa_{R,in}/2\pi = 5$ kHz.

While the triplet modes are clearly visible in the color maps of cavity spectrum, the lasers are only weakly coupled to them (by design), and our ability to accurately determine the resonance reflection dip and linewidth is limited. However, by averaging data from different membrane positions, we are able to produce values with sufficient accuracy for use in the theoretical model.

The linear couplings ($\omega'_{dis,L}$, $\omega'_{dis,R}$) and tunneling rate ($t$) determine the exact shape of the anti-crossings in the cavity spectra. To measure them, we again fit the Lorentzian peaks at each membrane position and record the center frequencies of each mode (see Fig. S4). The functional dependence of cavity resonant frequency on membrane position near the crossing is given by the eigenvalues of the $M$ matrix in equation (1) in the main text (a simple hyperbola, in the case of $\omega'_{dis,L} = \omega'_{dis,R}$). Instead of fitting to this, here we chose to fit the linear slopes far away from the crossing and find the tunneling rate $t$ by fitting the curves near the crossing to a quadratic (the second derivative of the eigenvalues of $M$ at $z_{dis} = 0$ nm relates $t$ to this quadratic coefficient). For the two-mode crossing in Fig. S3, we find $\omega'_{dis,L}/2\pi = 2.1$ MHz / nm, $\omega'_{dis,R}/2\pi = -1.8$ MHz / nm, and $t/2\pi = 4.6$ MHz.

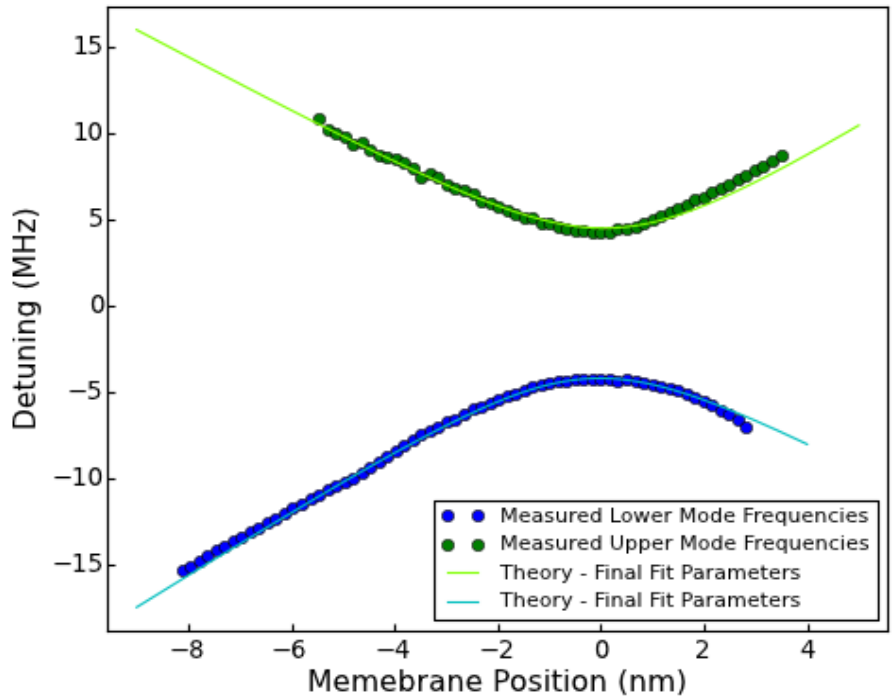

Figure S4: Plot of upper and lower mode resonance frequencies near the avoided crossing from Fig. S3, as found from Lorentzian fits. The solid lines are theory fits whose parameters are given in the text.

The final system parameter is the phase factor, $e^{i\phi}$. It is perhaps most instructive to think of $\phi$ as the phase acquired by a photon as it tunnels from one mode to the other. An alternate interpretation can be seen by removing this complex phase from the tunneling amplitude and instead having each mode couple to the laser drive with a different phase shift. It is physically correct to include both of these phases, but their effects on the model are equivalent, so we group them together as the complex phase of $t$. This phase shift affects the avoided crossing in measurable ways. The plots in Fig. S5 show the calculated effect of $\phi$ on the cavity spectrum near the crossing. We see clearly that when the optical modes hybridize, $\phi$ modifies the interference of the two modes and results in different relative coupling strengths. We determined $\phi$ for our system by measuring the relative coupling (comparing resonant reflection dips) at $z_{\text{dis}} = 0$ nm. We found $\phi = 1.6$ (approximately $\pi/2$, corresponding to equal dips at the quadratic point).

The case in which there are two avoided crossings between nearly-degenerate triplet modes and the singlet can be handled in almost exactly the same way as described above to measure $\omega'_{\text{dis,L}}$, $\omega'_{\text{dis,R}}$, $\phi$, and $t$ for each of the three modes. However, since the quadratic curvature is poorly resolved for the smallest crossing, we find $t$ for this crossing directly by measuring the gap between the two modes (instead of fitting the quadratic curvature). The result is $t_2/2\pi = 0.76$ MHz and the other results are listed in the Table S1. Note that the larger gap is denoted as the crossing $t_1$ between modes $L$ and $R_1$ and the smaller gap as the crossing $t_2$ between modes $L$ and $R_2$.

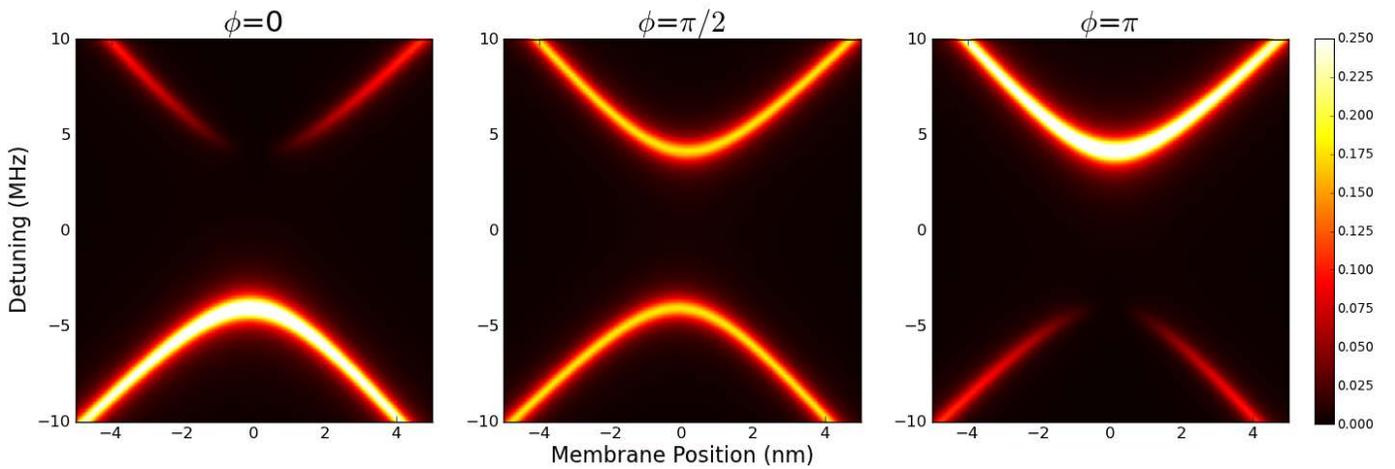

Figure S5: Cavity spectrum (calculated from theory) for three different values of the tunneling phase, $\phi$. Equally-coupled modes were used here to make the effect more visible.

## Fit results

We obtained most of the system parameters from the cavity reflection spectrum. The effective linear coupling, $\omega'_{\text{osc}}$, however, is not directly obtained from the spectrum. We include it as a fit parameter when fitting data measured with different membrane displacements and use the average value as a fixed system parameter for the final fit analysis. The average values of $\omega'_{\text{osc}}$ are listed in Table S1.

Table S1: System parameters used for the Figures in the main text

| System parameters | Figures in the main text | | | | |
|---|---|---|---|---|---|
| | Figure 2 | Figure 3 | Figure 4 | | |
| | | | I | II | III |
| $\omega''/2\pi^\dagger$ (MHz/nm$^2$) | 4.2 | 4.2 | 1.7 | 4.2 | 8.7 |
| $\omega'_{dis,L}/2\pi$ (MHz/nm) | 1.87 | 1.87 | 2.13 | 1.87 | 1.87 |
| $\omega'_{dis,R1}/2\pi$ (MHz/nm) | -1.77 | -1.77 | -1.82 | -1.77 | -1.77 |
| $\omega'_{dis,R2}/2\pi$ (MHz/nm) | -1.77 | -1.77 | N/A | -1.77 | -1.77 |
| $\omega'_{osc,L}/2\pi$ (MHz/nm) | 1.40 | 1.40 | 1.56 | 1.40 | fit parameter |
| $\omega'_{osc,R1}/2\pi$ (MHz/nm) | -1.46 | -1.46 | -1.66 | -1.46 | -1.46 |
| $\omega'_{osc,R2}/2\pi$ (MHz/nm) | -0.65 | -0.65 | N/A | -0.65 | fit parameter |
| $t_1/2\pi$ (MHz) | 1.57 | 1.57 | 4.57 | 1.57 | 1.57 |
| $t_2/2\pi$ (MHz) | 0.76 | 0.76 | N/A | 0.76 | 0.76 |
| $\kappa_L/2\pi$ (MHz) | 1.0 | 1.0 | 1.0 | 1.0 | 1.0 |
| $\kappa_{L,in}/2\pi$ (kHz) | 74 | 74 | 46.8 | 74 | 74 |
| $\kappa_{R1}/2\pi$ (MHz) | 1.3 | 1.3 | 1.3 | 1.3 | 1.3 |
| $\kappa_{R1,in}/2\pi$ (kHz) | 7 | 7 | 4.7 | 7 | 7 |
| $\kappa_{R2}/2\pi$ (MHz) | 1.3 | 1.3 | N/A | 1.3 | 1.3 |
| $\kappa_{R2,in}/2\pi$ (kHz) | 4 | 4 | N/A | 4 | 4 |
| $\phi_1$ | 1.9 | 1.9 | 1.6 | 1.9 | 1.9 |
| $\phi_2$ | 1.1 | 1.1 | N/A | 1.1 | 1.1 |
| $P_{in}$ ($\mu$W) | 40 | fit parameter | 80 | fit parameter | 80 |

$^\dagger$calculated value from $\omega'_{dis}$ and $t$

Table S2: Fit results used for the Figures in the main text

| Fit parameters | Figures in the main text | | | | |
|---|---|---|---|---|---|
| | Figure 2 | Figure 3 | Figure 4 | | |
| | | | I | II | III |
| $z_{dis}$ (nm) | See Fig. S6a | See Fig. S6b | $-0.42 \pm 0.05^\dagger$ | $0.36 \pm 0.01^\dagger$ | $-0.09 \pm 0.01^\dagger$ |
| $P_{in}$ ($\mu$W) | N/A | See Fig. S6c | N/A | $96.4 \pm 3.0^\dagger$ | N/A |
| $\omega'_{osc,L}/2\pi$ (MHz/nm) | N/A | N/A | N/A | N/A | $1.26 \pm 0.02^\dagger$ |
| $\omega'_{osc,R2}/2\pi$ (MHz/nm) | N/A | N/A | N/A | N/A | $-0.62 \pm 0.05^\dagger$ |

$^\dagger$statistical fit error

Control laser power $P_{in}$ is measured with a power meter at the entrance of the fiber prior to entering the cryostat. We consider ∼ 40% power loss through the fiber. Mechanical quality factor $Q$

≈ 100,000 is obtained from membrane ringdown time $\tau \approx 0.1$ s by measuring the decay of the membrane's vibration at 354.6 kHz after the application of a strong piezo drive. The effective mass of the membrane is calculated to be 43 ng based on its size and material properties (i.e. $Si_3N_4$ membrane of 1 × mm × 1mm × 50 nm). The system parameters and their values used for Fig. 2-4 in the paper are listed in Table S1 while Table S2 shows the fit results. Some of the results i.e. $z_{dis}$ and $P_{in}$ are compared with control values (Fig. S6a-c). Note that for the data analysis of 'I' in Fig. 4, two optical modes are considered: the singlet and one of the triplet modes. For the rest of the data, however, an additional triplet mode is included. This additional mode forms an avoided crossing nearby with the singlet mode (Fig. S7).

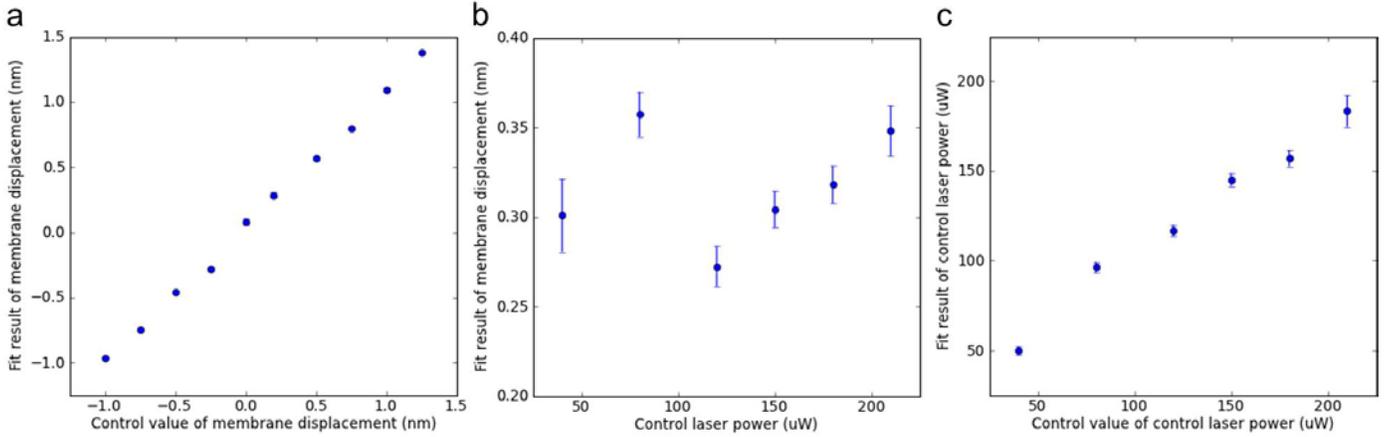

Figure S6: Fit results vs control values. a-c, the fit results used for Fig. 2 (a) and Fig. 3 (b-c). The fit results of membrane displacement $z_{dis}$ (a) and control laser power $P_{in}$ (c) are compared with their control values and show good agreement. The error bars denote statistical fit errors.

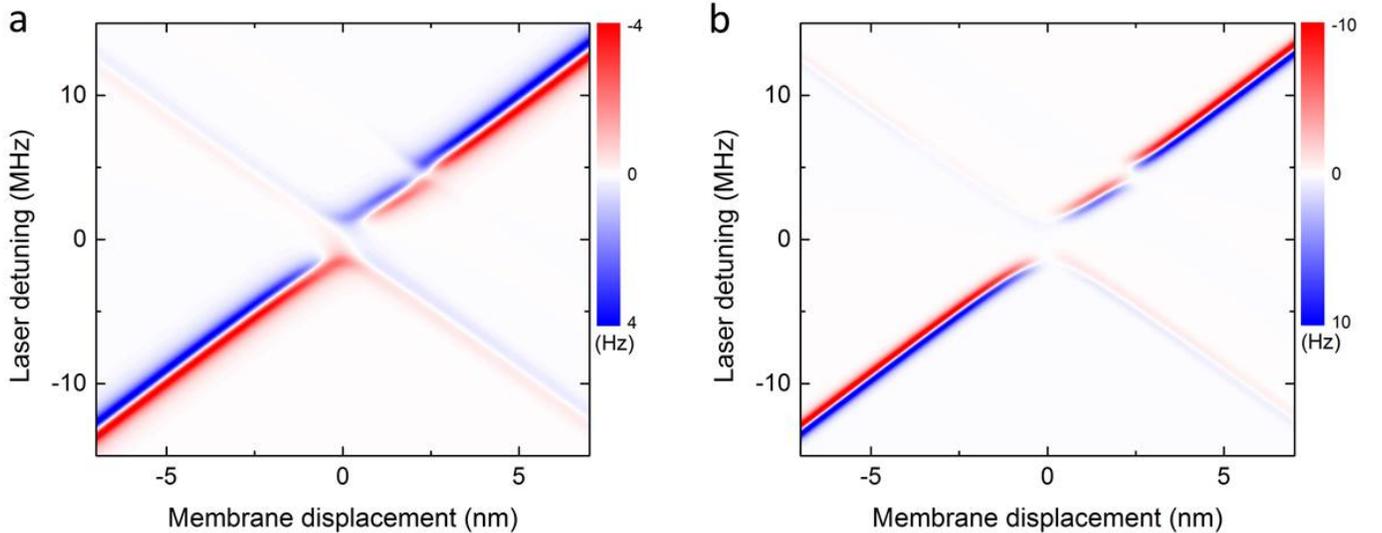

Figure S7: a-b, Calculated optical spring (a) and damping (b). The model includes one singlet mode and two of triplet modes.

# 3 Theory

Here, we outline our model for the optomechanical interactions arising from two coupled optical modes. We begin with a derivation of single (optical) mode optomechanics, then generalize this to two or more coupled optical modes.

## Optomechanics of a single optical mode

First, we review the derivation of optomechanics for a system with a single optical mode, in which the Hamiltonian is:

$$\widehat{\mathcal{H}} = \hbar(\omega_c + g_m \hat{z})\hat{a}^\dagger \hat{a} + \hbar \omega_m \hat{c}^\dagger \hat{c} + \widehat{\mathcal{H}}_{env} \tag{1}$$

The first term describes the optical cavity, while the second accounts for the mechanical motion. In this expression $\hat{a}$ and $\hat{c}$ are annihilation operators for the optical and mechanical modes, respectively, $\omega_c$ is bare cavity resonant frequency, $g_m$ is the linear optomechanical coupling for one phonon ($\frac{\partial \omega_c}{\partial z} z_{zpf}$ where $z_{zpf} = \sqrt{\hbar/2\omega_m m}$) and $\omega_m$ is the mechanical mode frequency. Mechanical displacement is expressed as $\hat{z} = \hat{c} + \hat{c}^\dagger$. Finally, $\widehat{\mathcal{H}}_{env}$ accounts for all coupling to the environment (decays and drives).

This Hamiltonian leads to the following equations of motion:

$$\dot{\hat{a}} = -(\kappa/2 + i\omega_c)\hat{a} - ig_m \hat{a}\hat{z} + \sqrt{\kappa_{in}}\hat{a}_{in} + \sqrt{\kappa_{vac}}\hat{a}_{vac} \tag{2}$$

$$\dot{\hat{c}} = -(\gamma_m/2 + i\omega_m)\hat{c} - ig_m \hat{a}^\dagger \hat{a} + \sqrt{\gamma_m}\hat{\eta} \tag{3}$$

Decay rates of the optical and mechanical modes are denoted as $\kappa$ and $\gamma_m$, respectively. $\kappa_{in}$ describes the coupling through the input port, which we use to drive the mode, while $\kappa_{vac} = \kappa - \kappa_{in}$ describes coupling to other dissipation channels. $\hat{a}_{in}$ and $\hat{a}_{vac}$ are drives through these two channels ($\hat{a}_{vac}$ is just vacuum noise, while $\hat{a}_{in}$ includes any external drives). Finally, $\hat{\eta}$ is the thermal drive for the mechanical mode.

For simplicity, we consider the (experimentally relevant) classical case, for which the equations of motion become

$$\dot{a} = -(\kappa/2 + i\omega_c)a - ig_m a z + \sqrt{\kappa_{in}} a_{in} \tag{4}$$

$$\dot{c} = -(\gamma_m/2 + i\omega_m)c - ig_m a^* a + \sqrt{\gamma_m}\eta \tag{5}$$

Next, we introduce an external coherent optical drive detuned by $\Delta$ from the cavity resonance: $a_{in}(t) = a_{in} e^{-i(\omega_c + \Delta)t}$, which (if we disregard mechanical motion and the negligible static displacement due to radiation pressure) creates a steady cavity optical field $a(t) = a_0 e^{-i(\omega_c + \Delta)t}$. The field's amplitude can be expressed as

$$a_0 = \frac{\sqrt{\kappa_{in}} a_{in}}{\kappa/2 - i\Delta} = \chi_c[0]\sqrt{\kappa_{in}} a_{in} \tag{6}$$

where $\chi_c[\omega]$ is the cavity susceptibility $\chi_c[\omega] = (\kappa/2 - i(\Delta + \omega))^{-1}$. We can now linearize our equations of motion around this coherent drive by writing $a(t) = (a_0 + d(t))e^{-i(\omega_c + \Delta)t}$ where $d(t) \ll a_0$:

$$\dot{d} = -(\kappa/2 - i\Delta)d - i\alpha z \tag{7}$$

$$\dot{c} = -(\gamma_m/2 + i\omega_m)c - i(\alpha^* d + d^*\alpha) + \sqrt{\gamma_m}\eta \tag{8}$$

Here, $\alpha = g_m a_0$ is the total optomechanical coupling. Taking the Fourier transform of these equations, we find

$$d[\omega] = -i\chi_c[\omega]\alpha z[\omega] \tag{9}$$

$$d^*[\omega] = +i\chi_c^*[-\omega]\alpha z[\omega] \tag{10}$$

$$c[\omega] = \chi_m[\omega](-i(\alpha^* d[\omega] + d^*[\omega]\alpha) + \sqrt{\gamma_m}\eta[\omega]) \tag{11}$$

$$c^*[\omega] = \chi_m^*[-\omega](i(\alpha^* d[\omega] + d^*[\omega]\alpha) + \sqrt{\gamma_m}\eta^*[\omega]) \tag{12}$$

Here we've introduced the mechanical susceptibility $\chi_m[\omega] = (\gamma_m/2 + i(\omega_m - \omega))^{-1}$.

Next, we substitute the expressions for $d[\omega]$, $d^*[\omega]$ into the mechanical equation of motion, multiply both of these equations by $(\chi_m[\omega]\chi_m^*[-\omega])^{-1}$ and add them together. Assuming that we're interested in frequencies $\omega \approx \omega_m$, and that $Q = \omega_m/\gamma_m \gg 1$, we can simplify $\chi_m^{-1}[-\omega] = \gamma_m/2 + i(\omega + \omega_m) \approx 2i\omega_m \gg \chi_m^{-1}[\omega]$. In the end, this allows us to obtain the solution

$$(\chi_m^{-1}[\omega] + i\Sigma[\omega])z[\omega] = \sqrt{\gamma_m}\eta[\omega] \tag{13}$$

From this, we see that the bare mechanical susceptibility $\chi_m^{-1}[\omega] = \gamma_m/2 + i(\omega_m - \omega)$ is modified by the self-energy term $\Sigma[\omega] = -i|\alpha|^2(\chi_c[\omega] - \chi_c^*[-\omega])$. Thus, changes in mechanical resonance frequency and linewidth can be expressed as $\delta\omega = \text{Re}(\Sigma[\omega_m])$, $\delta\gamma = -2\text{Im}(\Sigma[\omega_m])$.

## Optomechanics of coupled optical modes

Consider the case of two crossing optical modes, which we'll call left ($L$) and right ($R$). We will disregard mechanical motion for now, but still consider constant membrane displacement (as it provides a way to tune the resonant frequencies of the two optical modes). The Hamiltonian for this system is

$$\hat{\mathcal{H}}_0 = \hbar(\omega_0 + g_{0,L}z_0)\hat{a}_L^\dagger \hat{a}_L + \hbar(\omega_0 + g_{0,R}z_0)\hat{a}_R^\dagger \hat{a}_R + \hbar(te^{i\phi}\hat{a}_L^\dagger \hat{a}_R + te^{-i\phi}\hat{a}_R^\dagger \hat{a}_L) + \hat{\mathcal{H}}_{env} \tag{14}$$

The first two terms describe the behavior of the left and the right cavity modes. The optomechanical coupling rate of each mode to the membrane displacement is denoted as $g_{0,L}$ and $g_{0,R}$ (in the notation of the main text, these are equal to $\omega'_{dis,L}$ and $\omega'_{dis,R}$ multiplied by $z_{zpf}$). The membrane displacement, $z_0$, which is a unitless (normalized to $z_{zpf}$) parameter here, is chosen such that for $z_0 = 0$, the frequencies of both modes are equal to $\omega_0$. The third term describes tunneling between the two modes with rate $t$. Note that we have chosen to use a real coupling term $t$ and explicitly include a complex phase factor $e^{i\phi}$. This can be thought of as the phase acquired by a photon tunneling from one mode to another. In addition to this phase factor, we could have chosen to have each mode couple to the input drive with a different phase shift. These two effects, while both physical, have identical effects on the model, so here we choose to only include a tunneling phase.

It is natural now to introduce vector notation for these modes, denoting vectors with a single bar and matrices with a double bar. For later notational convenience, we will also move to a frame rotating at $\omega_0$, so that our mode crossing effectively occurs at $\omega_0 = 0$. Using the definitions

$$\bar{\hat{a}} = \begin{pmatrix} \hat{a}_L \\ \hat{a}_R \end{pmatrix} \tag{15}$$

$$\bar{\hat{a}}^\dagger = \begin{pmatrix} \hat{a}_L{}^\dagger & \hat{a}_R{}^\dagger \end{pmatrix} \tag{16}$$

$$\bar{\bar{\omega}}_c = \begin{pmatrix} 0 & te^{i\phi} \\ te^{-i\phi} & 0 \end{pmatrix} \tag{17}$$

$$\bar{\bar{g}}_0 = \begin{pmatrix} g_{0,L} & 0 \\ 0 & g_{0,R} \end{pmatrix} \tag{18}$$

the Hamiltonian simplifies to

$$\widehat{\mathcal{H}}_0 = \hbar \bar{\hat{a}}^\dagger (\bar{\bar{\omega}}_c + \bar{\bar{g}}_0 z_0)\bar{\hat{a}} + \widehat{\mathcal{H}}_{env} = \hbar \bar{\hat{a}}^\dagger \bar{\bar{\omega}}_c(z_0)\bar{\hat{a}} + \widehat{\mathcal{H}}_{env} \tag{19}$$

(DC optomechanical coupling is absorbed into $\bar{\bar{\omega}}_c(z_0) = \bar{\bar{\omega}}_c + \bar{\bar{g}}_0 z_0$).

We now switch to the classical description and express the equations of motion using the vector notation:

$$\dot{\bar{a}} = -(\bar{\bar{\kappa}}/2 + i\bar{\bar{\omega}}_c(z_0))\bar{a} + \overline{\sqrt{\kappa_{in}}}a_{in} \tag{20}$$

$$\bar{\bar{\kappa}} = \begin{pmatrix} \kappa_L & 0 \\ 0 & \kappa_R \end{pmatrix} \tag{21}$$

$$\overline{\sqrt{\kappa_{in}}} = \begin{pmatrix} \sqrt{\kappa_{L,in}} \\ \sqrt{\kappa_{R,in}} \end{pmatrix} \tag{22}$$

Here we account for the fact that the bare linewidths ($\kappa_L$ and $\kappa_R$) and input coupling rates ($\kappa_{L,in}$ and $\kappa_{R,in}$) can be different for the two modes. Since the same incident beam couples to both modes, $a_{in}$ is just a scalar, and the modes only differ in their coupling rates (as noted before, the phases of input coupling coefficients have been absorbed into our definitions of $a_L$ and $a_R$). Now we turn on an external drive detuned from the crossing point by $\Delta$, written (in the rotating frame) as $a_{in}(t) = a_{in}e^{-i\Delta t}$. This provides us with a steady state solution

$$\bar{a}(t) = \overline{a_0}e^{-i\Delta t} \tag{23}$$

$$\bar{a}_0 = (\bar{\bar{\kappa}}/2 + i(\bar{\bar{\omega}}_c(z_0) - \Delta))^{-1}\overline{\sqrt{\kappa_{in}}}a_{in} \tag{24}$$

$$= \bar{\bar{\chi}}_c[0]\overline{\sqrt{\kappa_{in}}}a_{in} \tag{25}$$

where scalars are assumed to be proportional to the identity matrix, i.e. $\Delta \equiv \bar{\bar{\Delta}} = \begin{pmatrix} \Delta & 0 \\ 0 & \Delta \end{pmatrix}$, and we've introduced the cavity susceptibility

$$\bar{\bar{\chi}}_c[\omega] = (\bar{\bar{\kappa}}/2 + i(\bar{\bar{\omega}}_c(z_0) - \Delta - \omega))^{-1} \tag{26}$$

Knowing this steady state solution we can, for example, find the reflected light amplitude as a function of $z_0$ and $\Delta$ (thus producing the sort of cavity spectra seen in Fig. 1d). The amplitudes of both cavity modes add coherently in the reflected light and we have

$$a_{refl} = a_{in} - \left(\sqrt{\kappa_{L,in}} a_{0,L} + \sqrt{\kappa_{R,in}} a_{0,R}\right) = a_{in} - \overline{\sqrt{\kappa_{in}}}^\dagger \bar{a}_0 \tag{27}$$

$$= a_{in}\left(1 - \overline{\sqrt{\kappa_{in}}}^\dagger \bar{\bar{\chi}}_c[0]\overline{\sqrt{\kappa_{in}}}\right) \tag{28}$$

Now we can add mechanical motion to our system. Depending on the overlap of the cavity modes with the particular mechanical mode, the optomechanical coupling will likely be reduced from the membrane displacement coupling ($g_{0,L/R}$). (For instance, if the cavity mode is centered near a nodal line of the mechanical mode, the resultant coupling will be significantly reduced.) We will denote the optomechanical coupling for an oscillating mode as

$$\bar{\bar{g}}_m = \begin{pmatrix} g_{m,L} & 0 \\ 0 & g_{m,R} \end{pmatrix} \tag{29}$$

Note that, as before, these coupling rates are normalized by $z_{zpf}$, so in the notation of the main text, $g_{m,L/R} = \omega'_{osc,L/R} z_{zpf}$. The mechanical motion will result in two additional terms in the Hamiltonian

$$\widehat{\mathcal{H}} = \hbar \bar{a}^\dagger \bar{\bar{g}}_m \bar{a} \hat{z} + \hbar \omega_m \hat{c}^\dagger \hat{c} + \widehat{\mathcal{H}}_0 \tag{30}$$

The first term accounts for the optomechanical coupling, while the second describes the mechanical motion. The equations of motion then transform into

$$\dot{\bar{a}} = -(\bar{\bar{\kappa}}/2 + i\bar{\bar{\omega}}_c(z_0))\bar{a} - i\bar{\bar{g}}_m \bar{a} z + \overline{\sqrt{\kappa_{in}}} a_{in} \tag{31}$$

$$\dot{c} = -(\gamma_m/2 + i\omega_m)c - i\bar{a}^\dagger \bar{\bar{g}}_m \bar{a} + \sqrt{\gamma_m}\eta \tag{32}$$

Using the steady state solution $\bar{a}_0$ from before we can, exactly as in the single mode case, linearize these equations:

$$\dot{\bar{d}} = -(\bar{\bar{\kappa}}/2 + i\bar{\bar{\omega}}_c(z_0) - i\Delta)\bar{d} - i\bar{\alpha}z \tag{33}$$

$$\dot{c} = -(\gamma_m/2 + i\omega_m)c - i(\bar{\alpha}^\dagger \bar{d} + \bar{d}^\dagger \bar{\alpha}) + \sqrt{\gamma_m}\eta \tag{34}$$

The total optomechanical coupling is now a vector $\bar{\alpha} = \bar{\bar{g}}_m \bar{a}_0$.

The derivation now follows the single-mode derivation nearly exactly, and we arrive at the final result:

$$\Sigma[\omega] = -i\bar{a}^\dagger(\bar{\bar{\chi}}_c[\omega] - \bar{\bar{\chi}}_c^\dagger[-\omega])\bar{a} \tag{35}$$

From which the optical spring and damping can be found via $\delta\omega = \text{Re}(\Sigma[\omega_m])$ and $\delta\gamma = -2\text{Im}(\Sigma[\omega_m])$.

Although slightly bulkier, this model of the optomechanics of multiple coupled optical modes is not significantly more complicated than the case of a single optical mode. The important feature of this model is that it universally describes a system that can exhibit both linear and quadratic coupling, depending on the static position of the membrane. Far away from the crossing, we can generate the canonical results for linear optical spring and damping, and as the membrane approaches the crossing point ($z_0 \to 0$) we see these linear effects vanish and the qualitatively different results of quadratic optomechanics arise.

The model discussed thus far is sufficient to predict the optomechanical effects from a single avoided crossing between two optical modes. In some of our experimental data, we deliberately introduced a second avoided crossing with a nearly-degenerate neighbor of one of the modes. We can easily extend our model to allow for multiple interacting modes by working with three-dimensional vector equations and introducing additional tunneling terms for the new mode. For instance:

$$\bar{a} = \begin{pmatrix} \hat{a}_L \\ \hat{a}_R \end{pmatrix} \to \begin{pmatrix} \hat{a}_L \\ \hat{a}_{R1} \\ \hat{a}_{R2} \end{pmatrix} \tag{36}$$

$$\overline{\sqrt{\kappa_{in}}} = \begin{pmatrix} \sqrt{\kappa_{L,in}} \\ \sqrt{\kappa_{R,in}} \end{pmatrix} \to \begin{pmatrix} \sqrt{\kappa_{L,in}} \\ \sqrt{\kappa_{R1,in}} \\ \sqrt{\kappa_{R2,in}} \end{pmatrix} \tag{37}$$

$$\bar{\bar{\kappa}} = \begin{pmatrix} \kappa_L & 0 \\ 0 & \kappa_R \end{pmatrix} \to \begin{pmatrix} \kappa_L & 0 & 0 \\ 0 & \kappa_{R1} & 0 \\ 0 & 0 & \kappa_{R2} \end{pmatrix} \tag{38}$$

$$\bar{\bar{g}}_0 = \begin{pmatrix} g_{0,L} & 0 \\ 0 & g_{0,R} \end{pmatrix} \to \begin{pmatrix} g_{0,L} & 0 & 0 \\ 0 & g_{0,R1} & 0 \\ 0 & 0 & g_{0,R2} \end{pmatrix} \tag{39}$$

$$\bar{\bar{g}}_m = \begin{pmatrix} g_{m,L} & 0 \\ 0 & g_{m,R} \end{pmatrix} \to \begin{pmatrix} g_{m,L} & 0 & 0 \\ 0 & g_{m,R1} & 0 \\ 0 & 0 & g_{m,R2} \end{pmatrix} \tag{39}$$

$$\bar{\bar{\omega}}_c = \begin{pmatrix} 0 & te^{i\phi} \\ te^{-i\phi} & 0 \end{pmatrix} \to \begin{pmatrix} 0 & t_1 e^{i\phi_1} & t_2 e^{i\phi_2} \\ t_1 e^{-i\phi_1} & 0 & 0 \\ t_2 e^{-i\phi_2} & 0 & \sigma \end{pmatrix} \tag{40}$$

where $\sigma$ is the frequency splitting between the nearly degenerate $R_1$ and $R_2$ modes and we've only allowed tunneling between each $R$ mode and the $L$ mode. Figures 1d and S7 show cavity spectra and optomechanical effects calculated using this three-mode theory.

14